\documentclass[sigconf,nonacm,10pt]{acmart}
\def\ARXIVVERSION{1}
\usepackage{tabularx}
\usepackage{dsfont}
\usepackage{multirow}
\usepackage{bm}

\newcommand{\figref}[1]{Figure \ref{#1}}
\newcommand{\tabref}[1]{Table \ref{#1}}

\usepackage{multirow}
\usepackage{enumitem}
\usepackage{adjustbox}
\usepackage{caption}
\usepackage{subcaption}
\usepackage{float}
\usepackage{placeins}
\newcommand{\compactdisplayskips}{%
  \setlength{\abovedisplayskip}{2pt}%
  \setlength{\belowdisplayskip}{3pt}%
  \setlength{\abovedisplayshortskip}{2pt}%
  \setlength{\belowdisplayshortskip}{3pt}%
}%
\usepackage{algorithm}
\usepackage{algpseudocode}
\usepackage[normalem]{ulem}
\usepackage{makecell}
\usepackage{array}
\usepackage{pifont}
\usepackage[nameinlink]{cleveref}
\crefname{section}{\S}{\S\S}
\crefname{subsection}{\S}{\S\S}
\Crefname{section}{\S}{\S\S}
\Crefname{subsection}{\S}{\S\S}

\useunder{\uline}{\ul}{}

\theoremstyle{plain}

\newtheorem{theorem*}{Theorem}

\usepackage{xspace}
\newcommand{\sys}{\textsc{Iapetus}\xspace}

\newcommand*{\RELEASE}{}

\ifdefined\RELEASE
  \newcommand{\ignore}[1]{}
  \newcommand{\fixme}[1]{}
  \newcommand{\TODO}[1]{}

  \newcommand{\cy}[1]{#1}

\else
  \newcommand{\ignore}[1]{}
  \newcommand{\fixme}[1]{{\textcolor{red}{[~FIXME:~#1~]}}}
  \newcommand{\TODO}[1]{{\textcolor{red}{TODO:~#1}}}

  \newcommand{\cy}[1]{{\textcolor{teal}{#1}}}

\fi

\AtBeginDocument{%
  \providecommand\BibTeX{{%
    \normalfont B\kern-0.5em{\scshape i\kern-0.25em b}\kern-0.8em\TeX}}}

\setcopyright{none}
\ifdefined\ARXIVVERSION
\else
  \copyrightyear{2026}
  \acmYear{2026}
  \acmConference[ACM Mobicom '26] {The 32nd Annual International Conference on Mobile Computing and Networking }{November, 2026}{Austin, Texas, USA.}
  \acmBooktitle{The 32nd Annual International Conference on Mobile Computing and Networking (ACM Mobicom '26), November, 2026 Austin, Texas, USA}
  \acmDOI{}
  \acmISBN{}
\fi

\makeatletter
\renewcommand{\footnotetextcopyrightpermission}[1]{}
\makeatother
\begin{document}

\title[Iapetus: Content-Aware Hierarchical Scheduling]{Iapetus: Content-Aware Hierarchical Scheduling for Collaborative ViT Inference in LEO Satellite Networks}

\author{\cy{Yan Chen, Yunxiang Zhang, Guanjun Jiang, and Haiquan Wang}}
\authornote{\cy{Corresponding author: Haiquan Wang (whq@buaa.edu.cn).}}
\affiliation{
  \institution{\cy{Beihang University}}
  \city{\cy{Beijing}}
  \country{\cy{China}}
}

\renewcommand{\shortauthors}{\cy{Chen et al.}}

\ifdefined\ARXIVVERSION
  \hypersetup{pdfauthor={Yan Chen, Yunxiang Zhang, Guanjun Jiang, Haiquan Wang}}
\fi

\begin{abstract}
Collaborative inference pools distributed resources to run compute-intensive Vision Transformers (ViTs) in satellite edge computing. Model partitioning enables such collaboration by assigning consecutive layer groups to different nodes, but the large volume of intermediate activation data incurs substantial transfer overhead that can erase its benefit. Token compression reduces downstream computation and activation transfer, but its quality impact depends on input content, model depth, and earlier pruning decisions, while layer offloading must adapt to time-varying contact and battery conditions. We present \sys, a content-aware hierarchical scheduler that screens constellation-wide options to retain a bounded candidate set, then refines each candidate into a complete token compression and layer offloading trajectory using quality prediction and joint planning. A unified objective balances per-task latency, energy, and quality loss against accumulated workload and battery pressures. We implement \sys on an NVIDIA Jetson AGX Orin hardware-in-the-loop testbed and use its validated execution model for constellation-scale trace replay across multiple ViT workloads and constellation settings. At \(5\)~tasks/s, \sys accomplishes 91.6\% of released tasks, 26.1 percentage points above MARATD3, the strongest baseline, while reducing mean latency and battery draw by 53.0\% and 70.8\%, respectively, and meeting quality targets.
\end{abstract}

\ifdefined\ARXIVVERSION
  \keywords{\cy{satellite edge computing, Vision Transformer, collaborative
  inference, trajectory planning, hierarchical scheduling}}
\fi

\maketitle

\section{Introduction}
\label{sec:introduction}

With the rise of satellite edge computing (SEC), low Earth orbit (LEO)
constellations are evolving beyond sensing and relay roles into platforms that
process Earth observation (EO) data in orbit~\cite{zhang2024energy,shi2025satellite}. Onboard
processing avoids downlinking every high-resolution image and enables timely
disaster, maritime, and environmental analysis~\cite{barmpoutis2020review,shi2025satellite}.
As onboard intelligence advances, Vision Transformers (ViTs)
are attractive backbones for global visual modeling~\cite{dosovitskiy2021image,le2024onboard},
with interest
in their deployment for in-orbit remote sensing~\cite{le2024onboard,baldwin2022leo1vit}.
However, bringing ViTs onboard shifts the bottleneck from downlink to
computation and energy: ViT-L/16 contains over 300M parameters and
requires approximately 190.7 GFLOPs per inference~\cite{xu2024devit}.
Such workloads consume scarce onboard compute and share the satellite's
power budget with communication and other spacecraft functions, while
inference during eclipse draws from a finite
battery whose sustained deep discharge can accelerate aging~\cite{liu2024orbit,fellner2003lithium}.
A sensing satellite may therefore capture a time-sensitive EO task that it
cannot finish locally or immediately offload to ground.

Collaborative inference helps bridge this resource gap by
exploiting compute beyond the sensing satellite. It can take two forms: full
offloading sends the raw input to a peer satellite or reachable ground
station~\cite{zhang2024energy,liu2024orbit}, while model partitioning assigns
consecutive layer groups to different nodes~\cite{chen2026hierarchical,zhang2026communication,deng2024daca}.
The latter enables finer-grained use of distributed compute, but shifts part
of the bottleneck to communication at partition boundaries. For
convolutional neural networks (CNNs), this
overhead is often modest because feature maps shrink with depth. ViTs behave
differently: ViT representations retain sequence length and embedding
dimension, making intermediate activations comparable to or larger
than the raw input~\cite{liu2025lvmscissor}. Transferring these activations
over an inter-satellite link (ISL) or ground-satellite link (GSL) can therefore
offset the gains from partitioning. In our measurement setup
(\cref{sec:motivation}), the fastest dense split takes 1.33~s, more than twice
the 0.64~s local inference latency. Token
compression mitigates both costs by
reducing subsequent computation and downstream activations, thereby lowering
computation, communication, and energy consumption~\cite{liang2026mercury,chen2024fastv}.

Fixed or decoupled pruning and layer offloading cannot adapt to heterogeneous inputs and system conditions.
EO images vary in texture, spatial structure, and redundancy, so the same pruning aggressiveness can cause different quality degradation across inputs. Under the same pruning setting (\cref{sec:motivation}), our measurements show no accuracy loss for Desert scenes but a 21.79~pp loss for Park scenes. A content-agnostic policy may retain unnecessary computation and traffic or prune too aggressively for the application target. Meanwhile, token compression and layer offloading are inherently coupled: a layer's compression decision changes all remaining computation and the payload of later transfers, which can determine whether those transfers fit within finite contact windows, while layer offloading determines where and when these costs occur~\cite{liang2026mercury}. We therefore formulate each scheduling decision as a complete content-aware token compression and layer offloading trajectory rather than independently selecting a compression ratio or a single partitioning point. Existing collaborative inference systems optimize compression and partitioning within predefined execution structures, but do not construct content-aware layer-wise trajectories under time-varying contacts~\cite{deng2024daca,chen2026hierarchical,zhang2026communication}.

\cy{Terrestrial mobile edge computing (MEC) systems typically assume a small
device group or a persistently reachable cloud path~\cite{deng2024daca,liang2026mercury}.
In LEO satellite networks, orbital motion instead exposes each task to finite ISL and
GSL windows and multiple satellite or ground candidates~\cite{lyu2023falcon,li2024short},
while workload and energy state continually change the available
resources for inference~\cite{zhang2024energy,liu2024orbit}. Applying this
joint adaptation across these candidates creates a constellation-scale online
planning problem. Fine-grained planning would require
collecting rapidly changing state across many satellites and evaluating
numerous layer-wise decisions, increasing coordination overhead and risking
stale information; purely local decisions can miss cross-node dependencies
in contacts, resources, and layer execution. Prior LEO systems similarly use
hierarchical or localized coordination to tame constellation-scale
dynamics~\cite{li2024short,chen2025partitioning}. This tension calls for
scalable coordination that limits global state collection while retaining the
local detail needed for feasible execution.}

These characteristics raise three key challenges.
\textbf{C1: Estimating cumulative quality effects along an evolving pruning trajectory.}
Pruning quality depends on input content, model depth, and the accepted
pruning prefix, but evaluating alternatives by inference is too costly online.
\textbf{C2: Constructing feasible token compression and layer offloading trajectories.}
Each pruning decision reshapes downstream computation and future transfer
payloads, while offloading determines where and when these costs occur; the
joint trajectory must satisfy finite contacts, battery limits, and deadlines.
\textbf{C3: Scaling trajectory planning across the constellation.}
Each
constellation-level candidate expands into a layer-wise joint
trajectory, so flat global planning would require rapidly
changing constellation state and a large combinatorial search.
Local planning alone, however, can miss dependencies needed
for feasible execution.

To address these challenges, we present \textsc{\sys}, an \textbf{I}n-orbit system for \textbf{A}daptive \textbf{P}runing and \textbf{E}xecution \textbf{T}rajectory planning with \textbf{U}nified \textbf{S}cheduling. \sys treats a complete content-aware token compression and layer offloading trajectory as its scheduling object for collaborative onboard ViT inference. Content-aware prediction and hybrid planning construct each trajectory, a common Lyapunov score compares complete plans, and hierarchical screening bounds the constellation-wide search. The leader associated with the selected plan then commits it once for fixed execution.

We implement \sys as a trace-driven SEC prototype calibrated with
measurements from two NVIDIA Jetson AGX Orin devices~\cite{liu2024orbit,cavu-aerospace-uk}
and validate its execution model on the corresponding hardware-in-the-loop
(HIL) testbed. Using this implementation, we evaluate \sys across
representative EO workloads and constellation settings. \cy{At $5$~tasks/s in
the default setting, \sys achieves a 91.60\% task accomplishment ratio (TAR).
Relative to MARATD3, the strongest baseline, this improves TAR by 26.1
percentage points (pp) and reduces mean latency and battery draw by 53.0\%
and 70.8\%, respectively, while meeting task-quality targets.}
The main contributions are as follows:
\begin{itemize}[leftmargin=*, nosep, topsep=0.25\baselineskip,
    partopsep=0pt, parsep=0pt, itemsep=0.25\baselineskip]

    \item We characterize why collaborative ViT inference
    requires planning token compression and layer offloading as a complete
    trajectory under LEO dynamics. Measurements show that dense
    activations can erase partitioning gains and that pruning sensitivity
    varies across input content.

    \item We develop \sys, a Lyapunov-guided hierarchical scheduler that
    combines content-aware quality prediction with joint compression and
    offloading across model layers. Its design screens a
    bounded set of constellation-level candidates before refining them into
    complete trajectories.

    \item \cy{We build a trace-driven prototype with an HIL testbed and evaluate it across three ViT backbones, two token compression strategies, and two constellation settings.
    Results demonstrate gains in task accomplishment, latency, and battery draw
    while meeting quality targets.}
\end{itemize}

\section{Motivation}
\label{sec:motivation}

This section uses three analyses to expose the design requirements for
collaborative ViT inference in SEC. We first examine the computation and
communication trade-off of model partitioning, then study how pruning tolerance
varies with image content, and finally show how time-varying conditions
change the preferred pruning and offloading decisions.

\noindent\textbf{Measurement methodology.}
\label{sec:motivation-methodology}
We fine tune ViT-L on AID~\cite{xia2017aid} and use it as the representative
workload, with per-layer execution measured on a Jetson AGX Orin. We compare
three execution primitives: local inference at the sensing satellite, full
peer offloading of the raw input, and model partitioning that transfers an
intermediate activation to a peer satellite. To study token compression, we use
ToMe~\cite{bolya2023token}, which progressively merges similar tokens without
retraining; \(r\) denotes the number of tokens reduced per Transformer block.\footnote{Local, Peer, Dense, and ToMe denote local-only inference,
full peer offloading, dense cut 1, and ToMe with \(r=8\) at cut 23,
respectively.}
We report latency and overall accuracy (OA).\footnote{Peer transfers the raw
input; Dense and ToMe transfer FP16 activations with 197 and 13 tokens
per model input, including the class token.}
For the dynamic analysis, we
\cy{release} tasks at a fixed rate of 5 tasks/s under time-varying system state; all
other settings follow \cref{subsec:evaluation-methodology}. These
measurements focus on onboard execution and inter-satellite collaboration,
while ground execution remains an optional tail when contact is available.

\begingroup
\begin{table}[t]
  \centering
  \caption{Representative ViT-L strategies at 0.1~Gbps under an accuracy-loss budget of 4~pp.}
  \label{tab:motivation-partition}
  \scriptsize
  \setlength{\tabcolsep}{2.5pt}
  \begin{tabular*}{\columnwidth}{@{\extracolsep{\fill}}lcccc@{}}
    \toprule
    Strategy & \makecell{Payload(MiB) \(\downarrow\)} & \makecell{Latency(ms) \(\downarrow\)} & \makecell{Battery draw(J) \(\downarrow\)} & \makecell{Accuracy loss(pp) \(\downarrow\)} \\
    \midrule
    Local & --    & 641.2  & 6.76  & 0.00 \\
    Peer  & 32.96 & 3023.8 & 32.17 & 0.00 \\
    Dense & 12.31 & 1326.3 & 17.22 & 0.00 \\
    ToMe  & 0.81  & \textbf{505.4} & \textbf{3.91} & 0.46 \\
    \bottomrule
  \end{tabular*}
\end{table}
\endgroup

\noindent\textbf{Intermediate activations limit ViT partitioning.}
\label{sec:motivation-pruning}
For each partitioned strategy in \tabref{tab:motivation-partition}, we scan all
internal cut points and report the fastest plan satisfying the accuracy budget.
At 0.1~Gbps, neither raw-input offloading nor dense partitioning outperforms
local inference: the best dense split still transfers 12.31~MiB and takes
1326.3~ms, compared with 641.2~ms locally. Dense ViT activations therefore
remain large enough for communication to offset the benefit of distributed
computation, consistent with prior observations on ViT intermediates
~\cite{xu2024devit}. Token compression changes this trade-off. The ToMe split
reduces the transferred activation to 0.81~MiB and completes in 505.4~ms while
drawing 3.91~J. Fewer tokens reduce both transfer volume and downstream
Transformer computation. Token compression is key to
beneficial ViT partitioning over constrained ISLs, but how aggressively tokens
can be removed depends on the input.

\begin{figure}[t]
  \centering
  \begin{subfigure}[t]{0.485\linewidth}
    \centering
    \includegraphics[width=\linewidth]{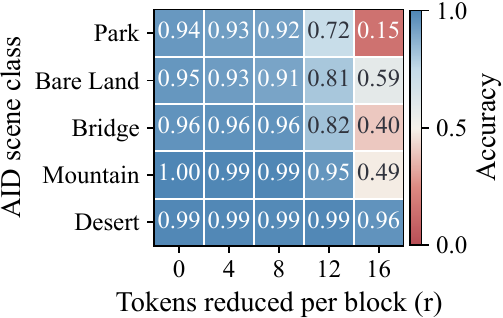}
    \caption{Accuracy loss across scene classes.}
    \label{fig:motivation-content-heatmap}
  \end{subfigure}
  \hfill
  \begin{subfigure}[t]{0.485\linewidth}
    \centering
    \includegraphics[width=\linewidth]{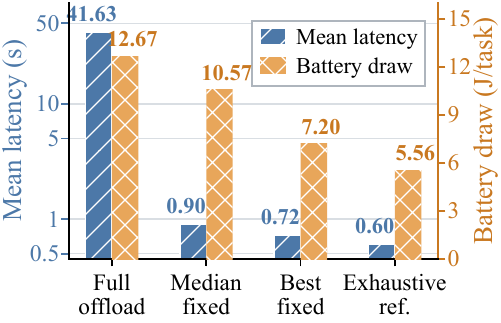}
    \caption{Exhaustive reference vs.\ fixed plans.}
    \label{fig:motivation-dynamic}
  \end{subfigure}
  \caption{Content sensitivity and limits of fixed plans.}
  \label{fig:motivation-overview}
  \Description{Two panels summarize the motivation experiments. Panel (a) is a heatmap of five representative AID scene classes across five ToMe settings, with Park, Bare Land, and Bridge degrading faster than Mountain and Desert. Panel (b) is a grouped bar chart comparing the mean latency and mean incremental battery draw of the exhaustive reference, the best fixed plan, the median fixed plan, and full peer offloading.}
\end{figure}

\noindent\textbf{Pruning sensitivity varies with image content.}
\label{sec:motivation-content}
Figure~\ref{fig:motivation-overview}(a) reports the OA of representative AID
scene classes under \(r\in\{0,4,8,12,16\}\), applied uniformly across
Transformer blocks. The same setting can produce sharply different accuracy
loss: at \(r=12\), Park loses 21.79~pp, whereas Desert shows no accuracy loss.
A fixed setting must therefore either remain conservative for tolerant images
or risk violating the quality target for sensitive ones. As further validated
in Section~\ref{subsec:design-validation}, model depth adds another dependency
because an early pruning action affects a longer suffix of the Transformer than
the same action near the output. Since scene labels and
counterfactual pruning outcomes are unavailable before inference, online
planning needs to estimate pruning sensitivity from raw image content while
accounting for model depth. Content awareness can determine how much pruning
an input tolerates, but the preferred execution location still depends on the
runtime system state.

\noindent\textbf{Static pruning and layer offloading are suboptimal.}
\label{sec:motivation-dynamic}
Figure~\ref{fig:motivation-overview}(b) compares the lowest-latency and median
fixed plans with an offline exhaustive reference over the measured
configurations. On a disjoint validation trace, we
retain plans with at least 99.9\% accomplishment, select the
lowest-latency plan (\(r=8\), cut 5) and a median-latency plan (dense, cut 4). \cy{Each fixed policy uses one configuration,}
whereas the reference selects the lowest-latency feasible plan for each
task. The
exhaustive reference and fixed policies accomplish approximately 99.99\% of tasks,
while full peer offloading accomplishes only 77.81\%. At comparable
accomplishment, the reference reduces mean latency and battery draw by
16.14\% and 22.88\%, respectively, relative to the best fixed plan. As
queueing, contacts, and available compute change, the preferred pruning
setting and partition point change with them. Pruning and layer offloading
must therefore be adapted jointly to the runtime state.

Together, these observations show that token compression enables ViT
partitioning, while pruning and offloading must adapt to input content and
runtime state. The exhaustive reference reveals this opportunity; \sys realizes
it through content-aware prediction and joint trajectory planning.

\FloatBarrier

\section{System Overview and Execution Abstraction}
\label{sec:system_model}
\label{sec:overview}

\sys organizes collaborative ViT scheduling by first selecting coarse
candidates and then constructing complete trajectories that jointly specify
token compression and layer offloading across model depth.

\subsection{System Architecture and Workflow}
\label{subsec:janussat-overview}

\begin{figure}[t]
  \centering
  \includegraphics[width=0.95\columnwidth]{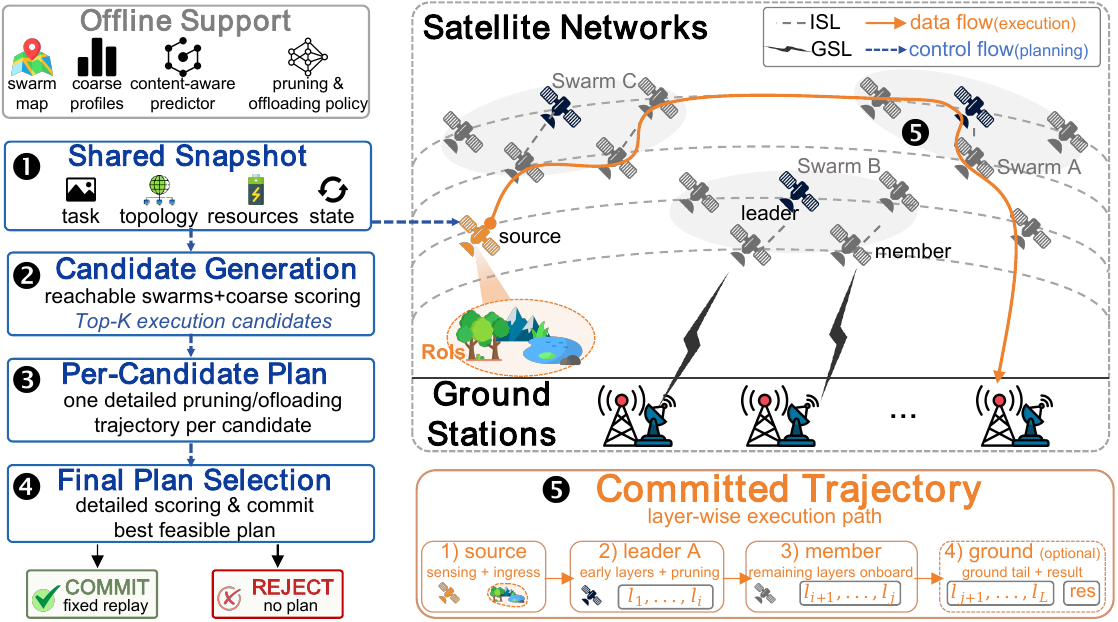}
  \caption{\cy{Overview of \sys.}}
  \label{fig:overview}
\end{figure}

\noindent\textbf{Architecture.}
Figure~\ref{fig:overview} illustrates the high-level architecture of \sys. When a satellite observes a designated region of interest (RoI), it becomes the sensing satellite for capturing EO task, which may then be processed collaboratively by other satellites or a reachable ground station. These nodes
communicate over time-varying ISL and GSL contacts. Satellites provide sensing,
computation, and communication resources under battery and energy-harvesting
constraints, while ground stations provide access to stronger ground-side
compute when visible. To organize this distributed execution, \sys uses
fixed, non-overlapping \emph{swarms} as local coordination units. Before
runtime, the nominal +Grid topology is partitioned using radius-one periodic
Lee coloring~\cite{klenze2018networking,golomb1970perfect}, providing stable
coordination domains with bounded membership, non-overlapping state, and
predictable control overhead. Each swarm contains
one leader and nearby members; the leader coordinates local planning, while
current contacts determine the members available for execution. Ground
stations remain optional endpoints outside the swarms.

\noindent\textbf{Workflow.}
Figure~\ref{fig:overview} summarizes the task-driven workflow of \sys.
The swarm map, cost models, content-aware predictor, and planning
policy are prepared offline and preloaded before runtime. For each task,
\ding{182} the sensing satellite combines the task and current system state
into a shared snapshot. \ding{183} It screens reachable swarms from this
snapshot and retains a bounded set of execution candidates
(\cref{subsec:lightweight-candidate}). \ding{184} The corresponding swarm
leaders concurrently construct complete token compression and layer offloading trajectories
and return the resulting plans (\cref{sec:inner}). \ding{185} The sensing
satellite applies detailed Lyapunov accounting and selects
the lowest-cost feasible plan, or rejects the task if none exists
(\cref{subsec:plan-selection}). \ding{186} The associated leader commits the
required resources and coordinates fixed execution across participating
satellites and an optional ground endpoint. The completed execution updates
the shared state for the next task.

\subsection{Decision and Execution Abstraction}
\label{sec:execution-abstraction}

Time is slotted with duration \(\Delta\). Let \(\mathcal S\) be the fixed
swarm set, \(\mathcal N_s\) the satellites in swarm \(s\), and \(L\) the
number of Transformer layers. Each task \(k\) released at slot \(t\) is
represented by
\((n_k,D_k^{\max},\rho_k^{\min},\boldsymbol\xi_k)\), denoting its sensing
satellite, deadline, quality target, and content descriptor, respectively.
Let \(\mathcal S_k(t)\) contain the swarms reachable from \(n_k\).
Let \(\mathcal M=\{\mathrm{sat},\mathrm{gnd}\}\) denote satellite-only and
ground-assisted execution, respectively. Each candidate
\(c=(s,m)\in\mathcal C_k(t)\subseteq
\mathcal S_k(t)\times\mathcal M\)
selects a swarm and execution mode. For
candidate \(c\), the layer planner constructs the trajectory
\(\tau_k(c)=(\boldsymbol\alpha_k(c),\boldsymbol\beta_k(c))\), whose
layer-\(\ell\) entries specify the fraction of incoming tokens removed and
the execution node, respectively. A node change transfers the current
activation; ground-assisted execution may enter one reachable ground endpoint,
after which ground is absorbing. Combining the candidate, trajectory, and
detailed resource and performance accounting forms the complete plan
\(q_k(c)\). A feasible plan satisfies the mode, contact, reservation, battery,
and deadline constraints at the time of commitment. For notational
simplicity, we omit the candidate
argument \(c\) when it is
clear from context.

The complete plan determines its realized latency, battery draw, and quality.
The pruning decisions up to layer \(\ell\) jointly determine the workload
\(W_{k,\ell}(\boldsymbol{\alpha}_{k,1:\ell})\) of that layer and the size of
the activation produced for a possible transfer. For a completed plan \(q_k\),
let
\(D_k(q_k)\) denote its total latency, including ingress, activation transfer,
waiting, and computation, and let \(E_k(q_k)\) denote its total onboard battery
draw from satellite computation and transmission. Transfer occurs only at an
offloading transition. Let \(\rho_k^{\mathrm d}>0\) be the quality achieved by
dense inference and \(\rho_k(q_k)\) the quality achieved by the completed
plan. The realized quality loss is
\(L_k(q_k)=[\rho_k^{\mathrm d}-\rho_k(q_k)]^+/\rho_k^{\mathrm d}\).
We use the following normalized cost to evaluate realized plan outcomes:
\par
{\footnotesize
\compactdisplayskips%
\begin{equation}
\Phi_k(q_k)
=\omega_D\frac{D_k(q_k)}{D_k^{\max}}
+\omega_E\frac{E_k(q_k)}{E_{\mathrm{ref}}}
+\omega_A L_k(q_k),
\label{eq:system-normalized-cost}
\end{equation}
}%
where \(E_{\mathrm{ref}}>0\) normalizes battery draw,
\(\omega_D,\omega_E,\omega_A\geq0\), and
\(\omega_D+\omega_E+\omega_A=1\). The completed execution meets the task's
quality target when \(\rho_k(q_k)\geq\rho_k^{\min}\).

\FloatBarrier

\section{Hierarchical Coordination}
\label{sec:outer}

\begin{figure}[t]
  \centering
  \includegraphics[width=0.95\columnwidth]{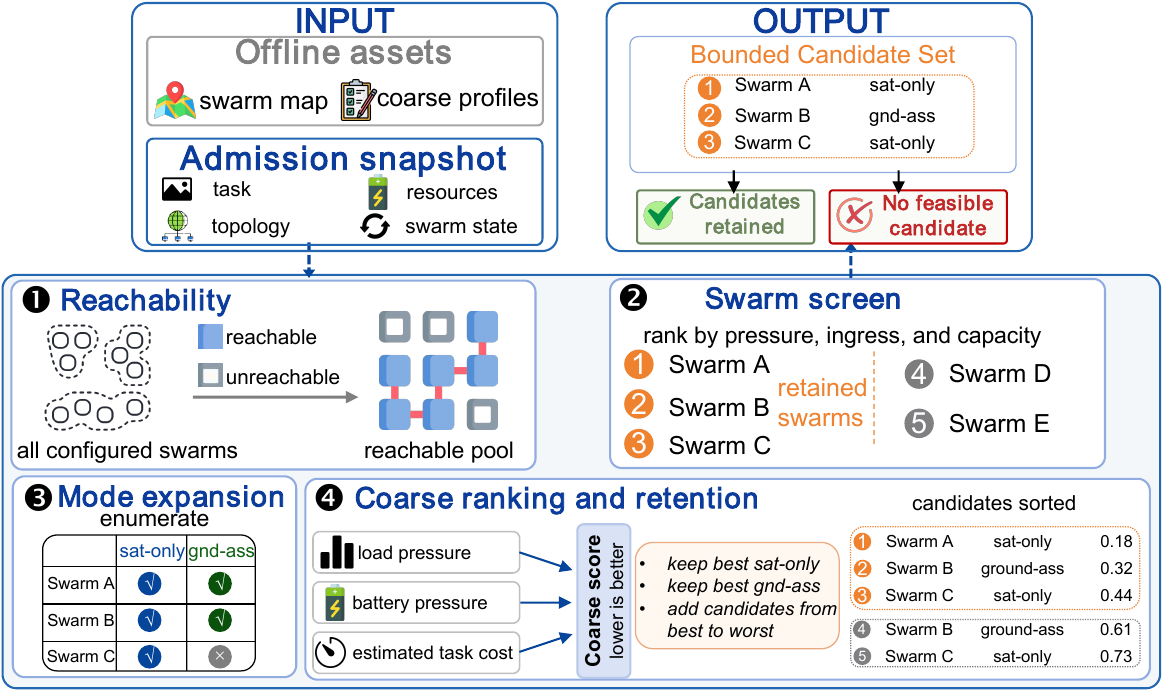}
  \caption{\cy{Bounded swarm--mode candidate generation.}}
  \label{fig:outer-scheduler-overview}
\end{figure}

\cref{sec:system_model} represents each scheduling decision as a coarse
candidate that is refined into a complete pruning and execution trajectory.
Exhaustively refining every reachable swarm--mode pair, however, would incur
substantial online planning overhead as the constellation grows. \sys avoids
this cost through Lyapunov-guided swarm screening, retaining only a bounded
set of promising candidates for detailed planning. As
Figure~\ref{fig:outer-scheduler-overview} shows, candidate generation proceeds
through reachability filtering (\ding{182}), swarm screening (\ding{183}),
mode expansion (\ding{184}), and candidate ranking and retention (\ding{185}).

\subsection{Lyapunov Score of a Complete Plan}
\label{subsec:lyapunov-score}

We first define the common plan score used as the reference for both coarse
candidate ranking and final plan selection. Selecting a swarm solely by its
execution cost can repeatedly concentrate workload and battery draw
on the same satellites. To capture these longer-term effects, \sys augments
each plan's execution cost with penalties that reflect the accumulated
workload and battery pressures. For a complete plan \(q_k\), we summarize its workload
and mean battery draw within swarm \(s\) as follows:
\par
{\footnotesize
\compactdisplayskips%
\begin{equation}
\begin{aligned}
W_{k,s}(q_k)
&=\sum_{\ell=1}^{L}
\mathbf 1[\beta_{k,\ell}\in\mathcal N_s]
W_{k,\ell}(\boldsymbol\alpha_{k,1:\ell}),\\
E_{k,s}(q_k)
&=\frac{\sum_{n\in\mathcal N_s}E_{k,n}(q_k)}{|\mathcal N_s|}.
\end{aligned}
\label{eq:swarm-task-accounting}
\end{equation}
}%
Ground execution contributes no satellite workload. We sum workload because
it consumes the aggregate service capacity of the swarm, whereas battery draw
is averaged to align with the mean battery state used to construct the battery
pressure. To characterize the battery state, let
\(B_n(t)\in[B_{\min},B_{\max}]\) denote the battery level of satellite \(n\),
and let
\(B_s(t)=\sum_{n\in\mathcal N_s}B_n(t)/|\mathcal N_s|\) denote the
swarm mean. Every feasible plan preserves the reserve \(B_{\min}\) at each
satellite.

To track accumulated workload pressure, let \(\mu_n(t)\) denote satellite
\(n\)'s compute capacity and
\(\mu_s(t)=\sum_{n\in\mathcal N_s}\mu_n(t)\) the aggregate capacity of swarm
\(s\). The swarm can therefore serve up to \(\Delta\mu_s(t)\) workload in one
slot. Let \(\mathcal A_t\) contain the tasks admitted in slot \(t\), and let
\(s_k\) denote the swarm of task \(k\)'s committed plan. The workload queue
evolves as
\par
{\footnotesize
\compactdisplayskips%
\begin{equation}
Q_s(t+1)=
[Q_s(t)-\Delta\mu_s(t)]^+
+\sum_{k\in\mathcal A_t}\mathbf 1[s_k=s]W_{k,s}(q_k).
\label{eq:queue-dynamics}
\end{equation}
}%
Here \(Q_s(t)\) is a virtual swarm workload queue that summarizes accumulated
workload pressure for scheduling, rather than the exact sum of node execution
queues. In parallel, \sys tracks persistent battery depletion through the
dimensionless virtual pressure
\par
{\footnotesize
\compactdisplayskips%
\begin{equation}
Z_s(t+1)=
\left[Z_s(t)+\frac{B_s(t)-B_s(t+1)}{B_{\mathrm{ref}}}\right]^+,
\label{eq:deficit-update}
\end{equation}
}%
where \(B_{\mathrm{ref}}>0\) normalizes battery variation. Net recharge lowers
\(Z_s(t)\), whereas sustained depletion raises it. Thus the reserve at each
satellite ensures instantaneous battery safety, while \(Z_s(t)\) discourages
persistent depletion over time.

With the workload and battery pressures defined, we follow Lyapunov
drift-plus-penalty control~\cite{neely2010stochastic} and collect the system
state as
\(\Theta(t)=\{Q_s(t),Z_s(t)\}_{s\in\mathcal S}\).
Using a quadratic Lyapunov function centered at the workload target
\(\theta_s\), and retaining the terms that depend on the current plan in the
standard drift bound, yields the following score for a plan \(q_k\) assigned
to swarm \(s\):
\par
{\footnotesize
\compactdisplayskips%
\begin{equation}
\Psi_{k,s}(q_k;t)=
\frac{[Q_s(t)-\theta_s]W_{k,s}(q_k)}{(Q_{\mathrm{ref}})^2}
+Z_s(t)\frac{E_{k,s}(q_k)}{B_{\mathrm{ref}}}
+V\Phi_k(q_k),
\label{eq:plan-dependent-phi}
\end{equation}
}%
where \(V>0\) controls the emphasis on the task cost,
\(Q_{\mathrm{ref}}>0\) normalizes workload, and \(\theta_s\) sets the desired
workload level, allowing the workload term to favor underloaded swarms. A lower
score is preferred. The first two terms penalize additional computation and
battery draw according to the accumulated swarm pressures, while the third
captures the execution cost of the current plan. This score provides a
ranking criterion for the bounded search: candidate generation uses coarse
estimates, whereas final selection uses detailed accounting.
Appendix~\ref{sec:appendix-lyapunov} provides the derivation.

\subsection{Coarse Candidate Generation}
\label{subsec:lightweight-candidate}

With the plan score defined, \sys progressively narrows the search using
increasingly detailed information: lightweight swarm summaries first bound the
coordination scope, after which mode-specific estimates rank the remaining
candidates under the same score.

\noindent\textbf{Reachability and swarm screening.}
\sys first removes swarms whose leaders have no usable ingress path from task
\(k\)'s sensing satellite under the current contact state
(\ding{182} in Figure~\ref{fig:outer-scheduler-overview}), yielding the
reachable set \(\mathcal S_k(t)\). \sys ranks each reachable swarm by the
normalized sum of workload pressure, battery pressure, ingress latency, and
estimated service time under its aggregate compute capacity. It retains up to
\(K_{\mathrm{swarm}}\) swarms in \(\mathcal S'_k(t)\) (\ding{183}). This
lightweight screening step limits the number of swarms considered before
execution modes are expanded for subsequent candidate construction.

\noindent\textbf{Mode expansion and coarse estimation.}
For each retained swarm \(s\in\mathcal S'_k(t)\), \sys instantiates the
satellite-only mode and, when a reachable ground tail exists, the
ground-assisted mode
(\ding{184} in Figure~\ref{fig:outer-scheduler-overview}). For each resulting
candidate \(c=(s,m)\), coarse models and measured lookup tables provide
nominal estimates of per-layer workload, activation payload, and quality.
The quality estimate uses the task descriptor, while detailed quality
prediction is deferred to trajectory planning.
These estimates are then combined with the current queue, compute, battery,
and contact state for each execution mode. For satellite-only mode,
\(\widehat W_k(c)\) covers full satellite execution; for ground-assisted mode,
it covers the satellite prefix. \(\widehat E_k(c)\) averages the estimated
satellite computation and transfer energy over the swarm, while
\(\widehat\Phi_k(c;t)\) combines the resulting latency and onboard energy with
predicted quality loss. For screening, an optimistic latency bound sums the estimated ingress,
aggregate swarm service, and optional ground-tail delay; detailed layer waiting
and transitions are deferred to trajectory planning.

\noindent\textbf{Candidate scoring and retention.}
\sys evaluates each candidate using the common Lyapunov score:
\par
{\footnotesize
\compactdisplayskips%
\begin{equation}
\widehat\Psi_k(c;t)=
\frac{[Q_s(t)-\theta_s]\widehat W_k(c)}{(Q_{\mathrm{ref}})^2}
+Z_s(t)\frac{\widehat E_k(c)}{B_{\mathrm{ref}}}
+V\widehat\Phi_k(c;t).
\label{eq:coarse-lyapunov-score}
\end{equation}
}%
Candidates that fail cheap necessary checks on contact availability, an
optimistic latency bound, or battery reserve are discarded before ranking.
Because coarse estimates may mis-rank execution modes, \sys first retains the
lowest-scoring surviving candidate from each available mode when the candidate
budget permits. It fills the remaining positions in global score order,
subject to
\(|\mathcal C_k(t)|\le K_{\mathrm{cand}}\)
(\ding{185} in Figure~\ref{fig:outer-scheduler-overview}). If
\(\mathcal C_k(t)=\varnothing\), the task is rejected; otherwise, every
candidate proceeds to trajectory planning in \cref{sec:inner}.

\FloatBarrier

\vspace{-0.25cm}
\section{Joint Trajectory Planning}
\label{sec:inner}

\begin{figure}[t]
  \centering
  \includegraphics[width=0.95\linewidth]{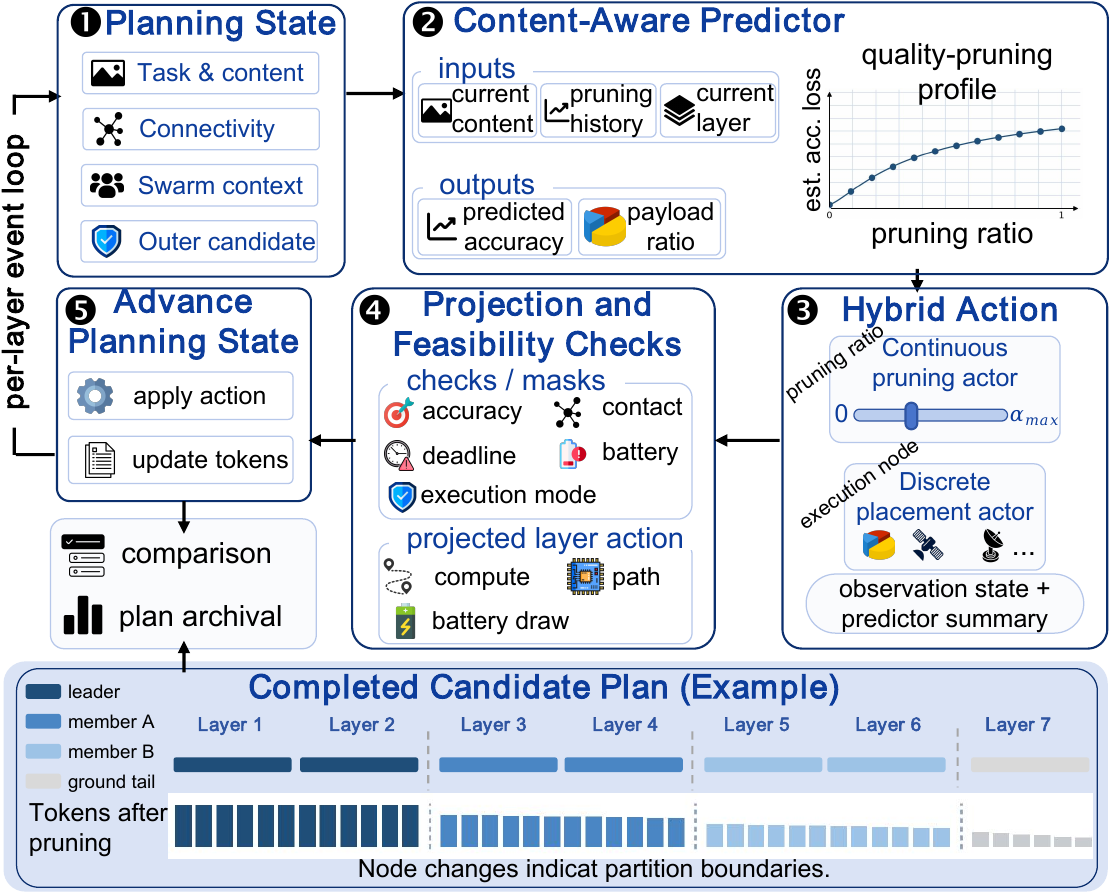}
  \vspace{2pt}
  \caption{\cy{Per-candidate sequential layer planning of joint pruning and offloading trajectories.}}
  \label{fig:inner-controller-overview}
\end{figure}

The bounded candidates from \cref{sec:outer} identify where detailed
planning is worthwhile but leave layer-wise execution unspecified. As
Figure~\ref{fig:inner-controller-overview} shows, the corresponding swarm
leaders concurrently construct a trajectory for each candidate,
jointly selecting pruning and layer offloading because each decision changes
subsequent computation, activation transfer, and resource availability. The
trajectory and its detailed accounting form a complete plan for final
selection.

\vspace{-0.5cm}
\subsection{Content-Aware Pruning Prediction}
\label{subsec:predictor}

The effect of token pruning depends on image content, model depth, and all
earlier pruning decisions. The same pruning control can preserve quality after
a mild prefix but remove critical information after an aggressive one. Executing the
unfinished ViT for every prospective pruning decision is too expensive during
planning,
so \sys uses a content-aware pruning predictor to estimate cumulative quality
effects across the evolving pruning trajectory.

\noindent\textbf{Prediction workflow.}
For task \(k\), a 16-dimensional descriptor \(\boldsymbol\xi_k\) summarizes
color, luminance, gradient, and texture statistics of its input. For each
prospective pruning control, the predictor combines this descriptor with the
normalized layer index and projected pruning control \(\alpha_{k,\ell}\) to
estimate the content- and depth-dependent final quality under dense
continuation. It then
uses the accepted prefix \(\boldsymbol\alpha_{k,1:\ell-1}\) to account for
interactions with earlier pruning. The prefix is encoded as an \(L\)-entry
layer-indexed vector, with future positions filled by the no-pruning control.
Denoting the complete predictor by \(f_\omega\), its interface is
\par
{\footnotesize
\compactdisplayskips%
\begin{equation}
\widehat\rho_{k,\ell}
=f_\omega\!\left(\boldsymbol\xi_k,\frac{\ell-1}{L-1},
\boldsymbol\alpha_{k,1:\ell-1},\alpha_{k,\ell}\right).
\label{eq:predictor-outputs}
\end{equation}
}%

Here \(\widehat\rho_{k,\ell}\) estimates final task quality if the
prospective control is accepted and the remaining layers continue without
further pruning. We use \(f_\omega\) to include a one-sided residual margin
calibrated on held-out validation histories, and \(\widehat\rho_{k,\ell}\)
denotes the resulting conservative estimate used for planning. At runtime,
before a layer decision is finalized, the leader queries the predictor for the
projected pruning control and carries the result into the next planning state;
a later query incorporates any newly accepted decision. Accordingly, the
resulting token count determines computation and activation payload directly
from the tensor shape.

\noindent\textbf{Training details.}
We train \(f_\omega\) from paired dense and pruned inference records that
cover both isolated pruning decisions and accumulated pruning histories.
Isolated records apply the queried control after a dense prefix. For each
history record, all recorded controls are replayed and the remaining layers
complete without further pruning, producing a final-quality label consistent
with the runtime query. Sampled histories encountered during offline trajectory
generation broaden the observed pruning patterns. A separate predictor is
trained for each supported pruning implementation. The default configuration
yields 1,134,000 isolated training records and
48,000 history-conditioned training records. The predictor uses
squared-error loss, a maximum tree depth of 6, and 120 boosting iterations.

\vspace{-0.3cm}
\subsection{Hybrid-Action Trajectory Planning}
\label{subsec:hybrid-policy}

With these pruning consequences available, \sys casts trajectory construction
for each candidate as a finite-horizon Markov decision process (MDP) with one
decision step per Transformer layer. Exhaustive search is combinatorial, while
decoupling pruning and layer offloading ignores their cross-layer coupling.
\sys therefore uses a hybrid-action policy to select the pruning ratio and
execution node jointly under deterministic feasibility constraints at each
layer.

\noindent\textbf{State and transition.}
At layer \(\ell\), state \(x_{k,\ell}\) summarizes the current trajectory,
including model progress, the activation holder and token state, remaining
local resources and contacts, and the fixed swarm pressures \(Q_s(t)\) and
\(Z_s(t)\). An accepted action updates this planning state and advances the
trajectory to the next layer for subsequent decision making.

\noindent\textbf{Hybrid action and feasibility.}
The action \(a_{k,\ell}=(\alpha_{k,\ell},\beta_{k,\ell})\) combines a
continuous pruning ratio with a discrete execution node assignment
(\ding{184} in Figure~\ref{fig:inner-controller-overview}). The continuous
proposal provides an ordered search space; \sys projects it onto a supported
native control before masking execution nodes that violate the resulting
contact, resource, battery, deadline, or mode constraints (\ding{185}). The
policy evaluates the remaining pruning and execution choices jointly.

\noindent\textbf{Layer reward.}
For a fixed candidate, the layer reward is the negative marginal increase in
the Lyapunov score of Eq.~\eqref{eq:plan-dependent-phi} caused by appending the
accepted action:
\par
{\footnotesize
\compactdisplayskips%
\begin{equation}
\Delta\Psi_{k,\ell}(c)=
\frac{Q_s(t)-\theta_s}{(Q_{\mathrm{ref}})^2}\Delta W_{k,\ell}(c)
+Z_s(t)\frac{\Delta E_{k,\ell}(c)}{B_{\mathrm{ref}}}
+V\Delta\Phi_{k,\ell}(c).
\label{eq:layer-lyapunov-reward}
\end{equation}
}%
The reward is \(r_{k,\ell}(c)=-\Delta\Psi_{k,\ell}(c)\). The first two terms
penalize additional workload and battery draw according to the current swarm
pressures, while the last captures the change in task execution cost. Because
these quantities are defined as marginal increments, they telescope along the
trajectory to the planning score of the completed plan. During training,
a constraint violation terminates the current trajectory with the corresponding
penalty; any trajectory that ends before model completion is excluded from plan
selection.

\noindent\textbf{Policy training.}
We train the trajectory policy offline on the ground using soft actor--critic
(SAC)~\cite{haarnoja2018soft}. The actor has two heads: one proposes the
continuous pruning ratio, and the other selects a discrete execution node. To
preserve their coupling, each twin critic conditions on state \(x\) and the
projected pruning action \(\alpha\), and outputs a value for every execution
node permitted by candidate \(c\):
\par\nopagebreak[4]
{\footnotesize
\compactdisplayskips%
\begin{equation}
\mathbf Q_i(x,\alpha)
=\left[Q_i(x,\alpha,\beta)\right]_{\beta\in\mathcal N(c)},
\qquad i\in\{1,2\}.
\end{equation}
}%
Here \(\mathcal N(c)\) is the node set permitted by the candidate's execution
mode. The planner applies the feasibility mask used at runtime to this
vector, allowing feasible node assignments to be evaluated under the same
pruning proposal while preserving their coupling with pruning-dependent
payload. Replay stores the projected pruning action so critic updates match
the executed transition. During training, terminal paired inference provides
realized quality feedback; during deployment, the content-aware prediction
supplies the quality signal used for trajectory construction. All critics and
optimizer state remain on the ground; deployment uses only the frozen actor
selected on validation data, and completed plans are ranked by Lyapunov
accounting rather than critic values.

\vspace{-0.3cm}
\subsection{Plan Selection and Commitment}
\label{subsec:plan-selection}

Coarse estimates are used only to bound the search space. Once trajectory
planning completes, \sys replaces them with detailed accounting of the
resulting computation, transfers, waiting, workload, and battery draw.
Incomplete or infeasible trajectories are discarded, and the sensing satellite
selects the complete plan with the lowest Lyapunov score:
\par
{\footnotesize
\compactdisplayskips%
\begin{equation}
q_k^\star
=\underset{c=(s,m)\in\mathcal C_k(t)}{\arg\min}\;
\Psi_{k,s}(q_k(c);t).
\label{eq:final-plan-selection}
\end{equation}
}%
If no feasible plan remains, the task is rejected.
Here \(\Psi_{k,s}\) is evaluated using complete plan accounting and
Eq.~\eqref{eq:plan-dependent-phi}.
The sensing satellite then sends \(q_k^\star\) to the associated leader.
The leader coordinates an atomic reservation request with participating nodes,
and execution begins once the request is accepted.
\sys executes the resulting layer-wise trajectory without
reinvoking the actor or switching to another candidate. This deferred
commitment ensures that all plans are evaluated before shared reservations
are updated.

\vspace{-0.25cm}
\section{Implementation}
\label{sec:implementation}

\noindent\textbf{Runtime and deployment.}
We implement the software prototype in Python using PyTorch and
scikit-learn. It consists of two parts. \textit{(i)} A trace-driven runtime
replays ISL, GSL, and sunlight intervals exported by Systems Tool Kit (STK)
and advances task, contact, compute, and battery events on a common timeline.
A routing module adapted from Falcon~\cite{lyu2023falcon} constructs alternative
ISL paths and ground egresses, while the resource manager maintains
processor and link ready times, battery state, and committed reservations.
\textit{(ii)} For distributed scheduling, the runtime instantiates sensing
satellite and swarm leader roles. The sensing satellite generates candidates, and
the corresponding swarm leaders construct trajectories independently
from local state. Leaders return complete plans without changing shared state;
after selection, the chosen leader commits the selected plan through the
resource manager, which atomically updates the affected compute and link
reservations. The coarse estimation models, content-aware predictor, frozen
policy, and active model checkpoint are preloaded before runtime. The
prototype comprises about \(30\mathrm{K}\) lines of code.

\noindent\textbf{Hardware profiling and HIL execution.}
We profile ViT execution on Jetson AGX Orin in 30-W and MAXN modes using NVIDIA \texttt{nvpmodel} and \texttt{jetson\_clocks}, recording per-layer latency, activation size, downstream FLOPs, and INA power.
Our HIL testbed comprises two Jetson AGX Orins and a desktop host.
One Orin represents the sensing satellite that originates each EO
task, while the other represents a collaborating peer; the desktop serves as
the ground-station endpoint. The two Orins communicate over dedicated Gigabit
Ethernet. We replay committed trajectories from the trace-driven runtime while
preserving their pruning controls, layer placements, and satellite/ground
transitions. Each assigned ViT segment and pruning action executes on the
corresponding physical device, and intermediate activations are serialized and
transferred over the physical link with pacing at the rates recorded in the trace.
For trajectories spanning multiple satellite hops, execution alternates
between the two Orins, such that every logical inter-satellite transition
incurs an actual device handoff and activation transfer. Ground-assisted
trajectories forward the current activation to the desktop for the remaining
ground-side execution.

\vspace{-0.25cm}
\section{Evaluation}
\label{sec:evaluation}

\cy{Our evaluation examines end-to-end performance under varying load, the
contributions of content-aware trajectory planning and bounded coordination,
robustness across system and workload settings, and physical execution
fidelity, runtime overhead, and scalability.}

\subsection{Evaluation Methodology}
\label{subsec:evaluation-methodology}
\label{sec:setup}

\noindent\textbf{Constellations, traces, and workload.}
We use subsets of the Starlink~\cite{starlink} and
OneWeb~\cite{weimerapplication} Walker constellations to cover distinct orbital
geometries. By default, the
controller operates on a \(24\)-plane Starlink subset. Each default test trace
spans \(6000\,\mathrm{s}\), approximately one orbit, including \(60\) minutes
of sunlight and \(35\) minutes of eclipse; \cref{subsec:robustness-generalization}
additionally evaluates OneWeb and a Starlink trace from another season.
Ground station locations follow StarPerf~\cite{starPerf}.
Tasks follow matched deterministic arrival traces. At rate \(\lambda\),
ten consecutive \(600\)-s phases contain \(6000\lambda\) releases, with one
sensing satellite active per phase. Unless otherwise stated, the
deadline is 30~s,
\cy{the quality target is 0.87 (about 8~pp below dense accuracy),}
\((K_{\mathrm{swarm}},K_{\mathrm{cand}})=(8,4)\), and
\((\omega_D,\omega_E,\omega_A)=(0.3,0.3,0.4)\).
Appendix \tabref{tab:appendix-system-parameters} and
\figref{fig:appendix-objective-weight-sensitivity} report the remaining
parameters and weight-sensitivity results.
We report means across seeds for all formal evaluation results.

\noindent\textbf{Models and datasets.}
To evaluate \sys across different model scales and computational loads, we use
ViT-L/16, ViT-H/14, and DINOv2-L/14~\cite{dosovitskiy2021image,oquab2024dinov2}.
We study scene
classification on two aerial image datasets. AID~\cite{xia2017aid} contains
10,000 \(600\times600\) RGB
images from 30 scene classes. RESISC45~\cite{cheng2017remote} contains 31,500
\(256\times256\) RGB images from 45 classes, with 700 images per class. Each
model uses a dataset-specific classification head and checkpoint, with inputs
resized to \(224\times224\). The image-disjoint training, validation, and test
splits contain 7,000/1,502/1,498 AID images and 22,500/4,500/4,500 RESISC45
images, respectively. Each task draws its content from the test split and
carries approximately \(0.27\,\mathrm{Gbit}\) of input before model resizing.
Unless otherwise stated, the main experiments use ViT-L on AID; the remaining
model--dataset combinations evaluate applicability across backbones and
scene distributions.

\noindent\textbf{Baselines.}
We compare \sys with five representative alternatives. \emph{(i)}
\textbf{GroundOnly} fully offloads each raw input through a reachable GSL,
representing ground-based execution. \emph{(ii)} \textbf{LocalDense} executes
the complete dense model on the sensing satellite without pruning or
offloading. \emph{(iii)} \textbf{Phoenix}~\cite{liu2024orbit} represents
sunlight-aware intersatellite task offloading without model partitioning.
\emph{(iv)} \textbf{MARATD3}~\cite{xiu2025computation} uses reinforcement
learning for joint offloading and resource allocation, treating each inference
request as an indivisible task. \emph{(v)} \textbf{SPS-AO}~\cite{zhang2026communication},
the closest baseline, partitions Transformer inference along an ordered
satellite chain and jointly selects activation compression.
All baselines use the same runtime, routing, and hardware profiles, with
identical orbital traces, arrivals, and image samples.

\noindent\textbf{Metrics.}
Task accomplishment ratio (TAR) is the fraction of all released tasks that
complete feasibly before their deadline and meet the task-quality target.
Rejected tasks, execution failures, deadline misses, and quality violations
are counted as unaccomplished. We also report end-to-end latency,
onboard battery draw per task, realized quality loss, and cost \(\Phi\) over accomplished tasks, using the definitions in
\cref{sec:execution-abstraction}. \cy{Reported latency covers the committed
execution path; scheduling overhead is evaluated separately in
\cref{subsec:overhead}.}\footnote{Latency comprises communication, waiting at the execution node after transfer completion, and computation. Communication includes contact and path waiting as well as data transmission. Battery draw includes onboard transmission and computation energy.}

\begin{figure}[t]
  \centering
  \includegraphics[width=0.92\columnwidth]{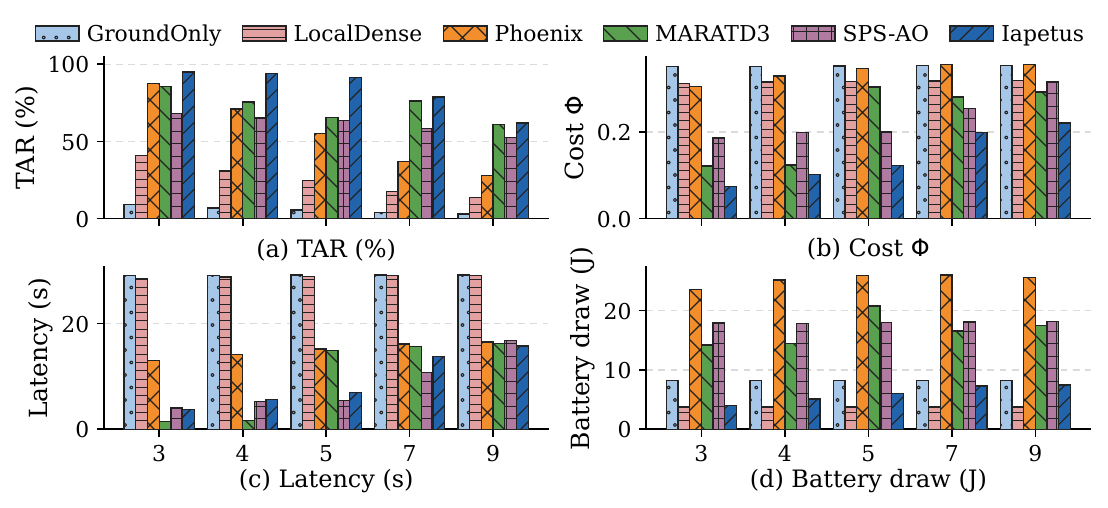}
  \caption{Performance under increasing load.}
  \label{fig:end-to-end-load}
\end{figure}

\begin{table}[t]
\centering
\caption{Latency and battery draw breakdown.}
\label{tab:end-to-end-summary}
\footnotesize
\setlength{\tabcolsep}{3.5pt}
\resizebox{\columnwidth}{!}{%
\begin{tabular}{lcccccc}
\hline
\multirow{2}{*}{Method} &
\multicolumn{3}{c}{Latency (s)} &
\multicolumn{2}{c}{Battery Draw (J)} &
\multirow{2}{*}{\mbox{Acc. loss (\%) \(\downarrow\)}} \\
\cmidrule(lr){2-4}\cmidrule(lr){5-6}
& Comm. & Wait. & Comp. & Comm. & Comp. & \\
\hline
GroundOnly & 29.13 & 0.00 & 0.02 & 8.21 & 0.00 & 0.00 \\
LocalDense & 0.00 & 28.05 & 0.85 & 0.00 & 3.77 & 0.00 \\
Phoenix & 13.77 & 0.61 & 0.84 & 22.04 & 3.92 & 0.00 \\
MARATD3 & 1.19 & 12.76 & 0.89 & 10.12 & 10.68 & 0.00 \\
SPS-AO & 0.75 & 3.88 & 0.77 & 6.50 & 11.47 & 2.99 \\
\sys & 6.79 & 0.04 & 0.15 & 5.38 & 0.69 & 1.74 \\
\hline
\end{tabular}}
\vspace{0.2cm}
\end{table}

\subsection{End-to-End Performance}
\label{subsec:end-to-end}

We sweep task arrival rates \(\{3,4,5,7,9\}\) in
Figure~\ref{fig:end-to-end-load}. At the representative rate of 5,
Table~\ref{tab:end-to-end-summary} further decomposes latency and onboard
battery draw and reports quality loss.

\noindent\textbf{TAR.}
Figure~\ref{fig:end-to-end-load}(a) shows that \sys consistently leads at every
evaluated load. At rates 3/5/9, it achieves 94.94/91.60/61.85\% TAR, compared
with 87.42/65.54/60.85\% for the strongest baseline at each rate. The baseline
trends explain this separation: intermittent GSL access and onboard waiting
reduce GroundOnly/LocalDense to 5.56/24.66\% at rate 5, while longer transfers
to distant sunlit satellites cause Phoenix to fall from 87.42\% at rate 3 to
28.00\% at rate 9. MARATD3/SPS-AO remain more competitive, but indivisible-task
placement and chain-constrained execution still limit contention relief. By
rate 9, aggregate link capacity constrains both \sys and MARATD3, narrowing
their gap to 1.00~pp. \sys therefore provides its clearest advantage before
system saturation by jointly adapting pruning, layer offloading, and swarm
selection to current content and system state.

\noindent\textbf{Cost.}
Figure~\ref{fig:end-to-end-load}(b) shows that \sys consistently obtains the
lowest cost at every evaluated load. At rate 5, its cost is 0.122,
39.1\%/59.8\% below SPS-AO/MARATD3, while accomplishing 91.60\% of tasks
compared with 63.60\%/65.54\%. Its lower cost does not arise from
averaging over fewer completed tasks, since \sys completes the largest
task set; instead, it better balances latency, battery draw, and
accuracy over that set.

\noindent\textbf{Latency.}
At rate 5, Figure~\ref{fig:end-to-end-load}(c) shows that \sys reduces mean
latency by 54.1\%/53.0\% relative to Phoenix/MARATD3. Table~\ref{tab:end-to-end-summary}
attributes their delays to 13.77~s of communication and 12.76~s of waiting,
while \sys limits waiting to 0.04~s. This keeps both transfer and queueing from
dominating execution. SPS-AO is faster at 5.40~s, but attains
only 63.60\% TAR, versus 6.98~s and 91.60\% for \sys.

\noindent\textbf{Battery draw.}
Figure~\ref{fig:end-to-end-load}(d) shows that \sys has the lowest battery draw
among collaborative methods at every load; its 6.07~J at rate 5 is 66.2\%
below SPS-AO. Table~\ref{tab:end-to-end-summary}
shows transmission/computation draws of 5.38/0.69~J for \sys and
6.50/11.47~J for SPS-AO. Token pruning therefore saves computation without
shifting energy to communication, while battery pressure avoids
energy-intensive plans.

Overall, \sys delivers the highest task accomplishment and lowest cost across
all evaluated rates while maintaining a favorable balance among latency,
battery draw, and accuracy.

\subsection{Component Evaluation}
\label{subsec:design-validation}

\noindent\textbf{C1: Content-aware predictor and token compression compatibility.}
We evaluate two representative token compression mechanisms with distinct
semantics. ToMe merges similar tokens~\cite{bolya2023token}, whereas TopK ranks
patch tokens by their activation \(\ell_2\) norm, retains the CLS token, and
discards the lowest-ranked patches. They represent token merging and token
pruning, respectively.
Table~\ref{tab:predictor-pruning-validation}(a) evaluates the three predictor
signals through matched counterfactual pairs that vary only image content,
layer placement, or the accepted pruning prefix. Full uses content, depth, and
history, whereas NoContent, NoDepth, and NoHistory remove the corresponding
signal. Each cell reports \emph{Full / Abla.}: pair accuracy measures ordering
agreement with realized final quality, regret measures the quality forgone by
the predicted choice, and the last two columns report unsafe selection and safe
acceptance at the 0.87 target. 
Table~\ref{tab:predictor-pruning-validation}(b)
then compares formal rate-5 systems using both mechanisms.

\begin{table}[t]
  \centering
  \caption{Predictor and token compression evaluation.}
  \label{tab:predictor-pruning-validation}
  \footnotesize
  \setlength{\tabcolsep}{2.5pt}
  \textit{(a) Matched-counterfactual predictor analysis.}\\[-2pt]
  \resizebox{\columnwidth}{!}{%
  \begin{tabular}{lllcccc}
    \hline
    Strategy & Signal & Pairs &
    Pair acc. (\%) \(\uparrow\) & Regret \(\downarrow\) &
    Unsafe (\%) \(\downarrow\) & Safe accept. (\%) \(\uparrow\) \\
    & & & \multicolumn{4}{c}{Full / Abla.} \\
    \hline
    \multirow{3}{*}{ToMe}
      & Content & 100 & 61.00 / 50.00 & 0.0256 / 0.0363 & 6.00 / 9.50 & 100.00 / 100.00 \\
      & Depth   & 90  & 65.56 / 50.00 & 0.0108 / 0.0156 & 0.00 / 0.00 & 100.00 / 100.00 \\
      & History & 96  & 83.33 / 50.00 & 0.0052 / 0.0156 & 0.00 / 0.00 & 100.00 / 100.00 \\
    \hline
    \multirow{3}{*}{TopK}
      & Content & 100 & 64.00 / 50.00 & 0.0200 / 0.0398 & 2.00 / 8.00 & 99.39 / 100.00 \\
      & Depth   & 55  & 65.45 / 50.00 & 0.0170 / 0.0261 & 7.27 / 9.09 & 100.00 / 90.77 \\
      & History & 76  & 76.32 / 50.00 & 0.0148 / 0.0238 & 7.89 / 10.53 & 97.59 / 96.39 \\
    \hline
  \end{tabular}}

  \setlength{\tabcolsep}{3pt}
  \textit{(b) End-to-end compatibility at rate 5.}\\[-2pt]
  \resizebox{0.92\columnwidth}{!}{%
  \begin{tabular}{lccccc}
    \hline
    Strategy & TAR (\%) \(\uparrow\) & $\Phi$ \(\downarrow\) &
    Latency (s) \(\downarrow\) & Battery draw (J) \(\downarrow\) &
    Acc. loss (\%) \(\downarrow\) \\
    \hline
    ToMe & 91.60 & 0.122 & 6.98 & 6.07 & 1.74 \\
    TopK & \cy{93.28} & \cy{0.139} & \cy{8.08} & \cy{7.27} & \cy{0.87} \\
    \hline
  \end{tabular}}
  \vspace{0.2cm}
\end{table}

Table~\ref{tab:predictor-pruning-validation}(a) shows that all three signals
matter across both compression mechanisms. Removing any one collapses pair
accuracy to the 50\% tie baseline; Full improves it by 11.00--33.33~pp and
reduces quality regret by 29.3--66.7\%. Content captures input-specific
sensitivity, depth identifies where pruning occurs, and history represents the
accumulated effect of earlier decisions. The unsafe-selection and safe-acceptance
results further show that these gains do not come from accepting greater
quality risk.

Table~\ref{tab:predictor-pruning-validation}(b) validates compatibility with
both merging-based and selection-based token compression. The complete
system achieves 91.60\% TAR with ToMe and \cy{93.28\%} with TopK while keeping
accuracy loss at 1.74\% and \cy{0.87\%}, respectively. Thus, the same planning
interface can incorporate distinct compression semantics without changing the
joint pruning and layer offloading workflow.

\noindent\textbf{C2: Joint pruning and layer offloading.}
Table~\ref{tab:joint-planning-ablation} isolates adaptive pruning,
joint pruning and layer offloading, and energy-aware coordination under the
rate-5 workload. FixedPrune uses one pruning ratio selected on validation data;
SinglePrune permits one pruning event; NoPrune disables pruning.
Decoupled optimizes pruning and layer offloading independently, while NoEnergy
removes the energy term.
Variants changing semantics use matched retraining. Results are
averaged across seeds and reported relative to \sys-ToMe in
Table~\ref{tab:predictor-pruning-validation}(b). TAR differences are in
pp; changes in \(\Phi\), latency, battery draw, and accuracy loss are
percentages.

\begin{table}[t]
\centering
\caption{Joint pruning and layer offloading ablations.}
\label{tab:joint-planning-ablation}
\scriptsize
\renewcommand{\arraystretch}{0.82}
\setlength{\tabcolsep}{2pt}

\begin{tabular*}{\columnwidth}{@{\extracolsep{\fill}}lccccc}
\toprule
Variant &
$\Delta$TAR $\uparrow$ &
$\Delta\Phi$ $\downarrow$ &
$\Delta$Lat. $\downarrow$ &
$\Delta$Batt. $\downarrow$ &
$\Delta$Acc. loss $\downarrow$ \\
\midrule
FixedPrune
& $-19.20$ & $+7.6$ & $-0.1$ & $-2.5$ & $+147.4$ \\
SinglePrune
& $+3.20$ & $+13.0$ & $+16.4$ & $+20.8$ & $-77.0$ \\
NoPrune
& $+2.88$ & $+12.1$ & $+16.1$ & $+22.4$ & $-100.0$ \\
Decoupled
& $-15.15$ & $+6.0$ & $+0.0$ & $-2.5$ & $+118.1$ \\
NoEnergy
& $+0.46$ & $+4.4$ & $-15.8$ & $+45.7$ & $-77.8$ \\
\bottomrule
\end{tabular*}
\vspace{0.2cm}
\end{table}

FixedPrune yields only marginal latency and battery savings, while losing
19.20~pp of TAR and increasing accuracy loss by 147.4\%. SinglePrune and
NoPrune instead raise TAR by 3.20 and 2.88~pp and reduce accuracy loss by
77.0\% and 100.0\%, but increase $\Phi$ by 13.0\% and 12.1\%, latency by
16.4\% and 16.1\%, and battery draw by 20.8\% and 22.4\%, respectively. Thus,
neither a fixed pruning ratio nor restricted or absent multi-layer adaptation
achieves Full's balance between accomplishment, quality, and resource use.
Decoupled lowers battery draw by 2.5\%, but loses 15.15~pp of TAR, raises
accuracy loss by 118.1\%, and increases $\Phi$ by 6.0\%. These losses in
accomplishment and quality confirm the need to optimize pruning and layer
offloading jointly. NoEnergy exposes the role of the energy-aware objective
most clearly: it lowers latency by 15.8\%, raises TAR by 0.46~pp, and reduces
accuracy loss by 77.8\%, but increases $\Phi$ by 4.4\% and battery draw by
45.7\%. Energy awareness prevents these aggressive execution choices while
Full maintains the best overall balance across accomplishment, quality,
latency, and battery draw.

\noindent\textbf{C3: Candidate reduction and Lyapunov alignment.}
\label{subsec:candidate-reduction}
Table~\ref{tab:candidate-reduction} evaluates candidate budgets, deferred
detailed selection, and Lyapunov pressure terms.
Table~\ref{tab:candidate-reduction}(a) compares the default \((8,4)\) design
with smaller and larger budgets; CoarseTop1 passes only the top coarse candidate
to complete planning, while NoPressure removes \(Q_s\) and \(Z_s\) with matched
retraining. Table~\ref{tab:candidate-reduction}(b) compares exhaustive detailed
scoring on 357 feasible instances from 360 phase-stratified samples.\footnote{We sample 120 tasks per seed. One task at the horizon boundary in each
seed admits no feasible plan, so its exhaustive optimum and score gap are
undefined; we exclude it from the task set.}
Recall
measures whether the exhaustive minimum-score candidate is retained, while the
P95 gap measures the resulting normalized score loss.
Table~\ref{tab:candidate-reduction}(a) reports changes relative to \sys-ToMe
result in Table~\ref{tab:predictor-pruning-validation}(b).

\begin{table}[t]
\centering
\caption{Candidate reduction and Lyapunov alignment.}
\label{tab:candidate-reduction}
\footnotesize
\setlength{\tabcolsep}{3.5pt}
\textit{(a) End-to-end changes relative to \sys-ToMe.}\\[-2pt]
\resizebox{\columnwidth}{!}{%
\begin{tabular}{lccccc}
\hline
Policy / $(K_s,K_c)$ & $\Delta$TAR (pp) \(\uparrow\) &
$\Delta\Phi$ (\%) \(\downarrow\) & $\Delta$Lat. (\%) \(\downarrow\) &
$\Delta$Batt. (\%) \(\downarrow\) & $\Delta$Acc. loss (\%) \(\downarrow\) \\
\hline
Full / (4,1) & $+0.61$ & $+7.5$ & $+16.1$ & $-5.2$ & $+5.3$ \\
Full / (16,8) & $-3.38$ & $+3.5$ & $+1.7$ & $+7.4$ & $-6.5$ \\
CoarseTop1 / (8,1) & $+0.70$ & $+8.3$ & $+18.1$ & $-6.2$ & $+6.1$ \\
NoPressure / (8,4) & $-0.68$ & $-3.1$ & $-11.0$ & $+11.3$ & $-24.7$ \\
\hline
\end{tabular}}
\textit{(b) Retention quality against exhaustive detailed scoring (\(N=357\)).}\\[-2pt]
{\setlength{\tabcolsep}{5pt}%
\begin{tabular}{@{}lcccc@{}}
\hline
Metric / $(K_s,K_c)$ & (4,1) & (8,1) & (8,4) & (16,8) \\
\hline
Recall (\%) \(\uparrow\) & 86.6 & 89.1 & 97.5 & 100.0 \\
P95 gap (\%) \(\downarrow\) & 24.76 & 24.76 & 1.66 & 0.00 \\
\hline
\end{tabular}}
\vspace{0.2cm}
\end{table}

Table~\ref{tab:candidate-reduction}(a) shows that the default \((8,4)\) budget
provides a balanced choice.
CoarseTop1 raises TAR, $\Phi$, and latency, lowers battery draw, and raises
accuracy loss. NoPressure lowers TAR, $\Phi$, and latency but raises battery
draw. The design balances these tradeoffs,
supporting deferred detailed selection and pressure-aware coordination.

Table~\ref{tab:candidate-reduction}(b) quantifies the quality of candidate
retention. With \((8,1)\),
the plan with the minimum score is retained for 89.1\% of tasks and the P95
normalized gap is 24.76\%; \((8,4)\) raises recall to 97.5\% and reduces the
gap to 1.66\%.
Increasing the budget to \((16,8)\) reaches 100\% recall and zero gap, showing
that a small set of complete plans captures most of the benefit of exhaustive
search. Higher recall does not imply monotonic end-to-end gains,
because committed plans alter workload and battery states.
\cref{subsec:overhead} complements this analysis of selection quality by
measuring how candidate budgets affect planning latency and summary traffic.

\vspace{-0.2cm}
\subsection{Robustness and Applicability}
\label{subsec:robustness-generalization}

\begin{figure}[t]
  \centering
  \includegraphics[width=\columnwidth]{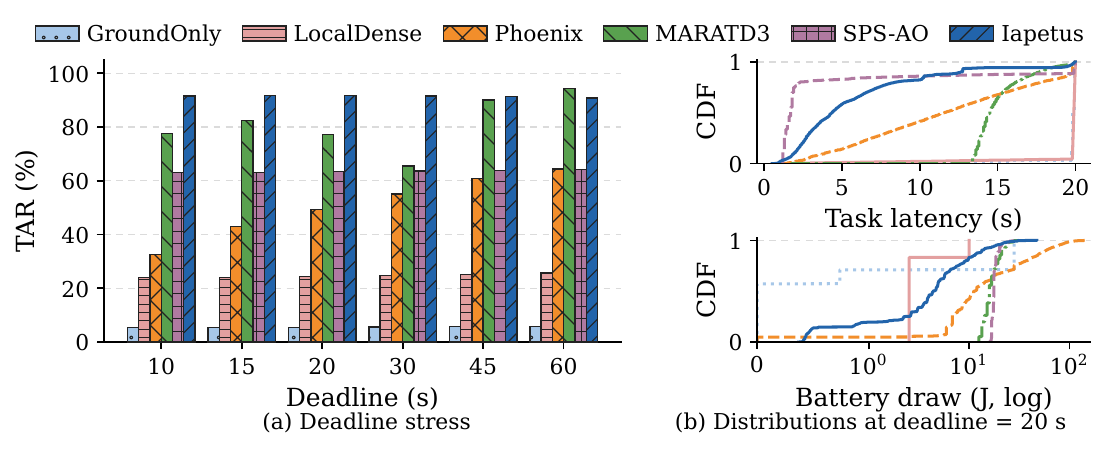}
  \caption{Deadline sensitivity.}
  \label{fig:deadline-sensitivity}
\end{figure}

\noindent\textbf{Impact of deadline.}
Figure~\ref{fig:deadline-sensitivity} varies the service deadline from 10 to
60~s at 5 tasks/s and reports completed-task latency and battery-draw CDFs at
20~s. \sys maintains a TAR of 90.88--91.67\% throughout the sweep and leads
every alternative through the 45-s deadline. Under the tight and moderate
deadlines of 10--30~s, it leads the next-best method, MARATD3, by
9.23--26.05~pp; at 45~s the margin remains 1.14~pp. MARATD3 benefits more from
additional slack and overtakes \sys by 3.54~pp only at 60~s, showing that
\sys's advantage is strongest when timeliness is binding. At 20~s, \sys has
mean/P95 latency of 5.87/19.30~s and mean/P95 battery draw of
6.00/18.46~J. Phoenix/GroundOnly report P95 latency of 19.82/19.92~s and P95
battery draw of 77.12/28.00~J. LocalDense draws only 3.77/9.97~J at mean/P95,
but its 19.46/19.99~s latency is concentrated near the deadline and its TAR
remains only 23.95--25.75\% across the sweep. Thus, \sys's high TAR under tight deadlines does not come at the cost of heavier
latency or battery tails.

\begin{figure}[t]
  \centering
  \scalebox{1}[1.1]{%
    \includegraphics[width=\linewidth]{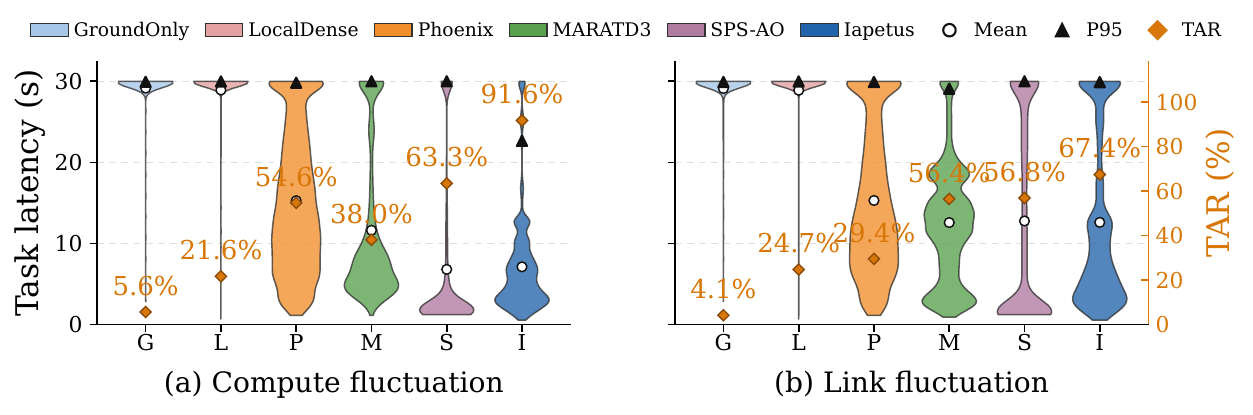}
  }
  \caption{Robustness to resource fluctuations.}
  \label{fig:resource-jitter}
  \vspace{0.2cm}
\end{figure}

\noindent\textbf{Impact of fluctuations.}
We evaluate robustness in the Starlink constellation at 5~tasks/s under
fluctuations in compute capacity and link bandwidth. Compute capacity is scaled
by $1+\min(0,X)$, where $X\sim\mathcal{N}(0,0.3^2)$, allowing only degradation.
Link fluctuations follow a Gilbert--Elliott model with good-to-bad and
bad-to-good transition probabilities of $0.15$ and $0.25$, respectively, where
the bad state reduces effective bandwidth by $80\%$.
Figure~\ref{fig:resource-jitter} reports TAR together with the per-task
latency distribution over accomplished tasks.

Under compute fluctuations, Figure~\ref{fig:resource-jitter}(a) shows that
\sys reaches 91.64\% TAR, 28.29~pp above SPS-AO, while keeping P95 latency at
22.65~s. Under the more severe link fluctuations in
Figure~\ref{fig:resource-jitter}(b), \sys reaches 67.35\% TAR,
10.54/10.97~pp above SPS-AO/MARATD3, despite comparable latency among
accomplished tasks. LocalDense reaches only 21.58/24.66\% TAR under
compute/link fluctuations, with P95 latency remaining at 29.99~s in both
settings. The admission statistics in Appendix
Table~\ref{tab:appendix-link-admission} explain this advantage:
\sys/SPS-AO/MARATD3 admit 72.0/78.1/56.4\% of released tasks and fulfill
93.5/72.7/100.0\% of their admissions, respectively. Thus, \sys maintains a
tighter latency tail under compute degradation and balances admission against
available resources when links fluctuate.  

\noindent\textbf{Impact of season.}
Across the 100-minute trace, the June setting in
Table~\ref{tab:transfer-summary}(a) changes the sunlight state at 16.34\% of
matched satellite-time points relative to October; the mean illumination shift
across individual planes ranges from $-5.01$ to $+1.74$~pp. With all policies
frozen, \sys sustains a 91.34\% TAR, only 0.26~pp below its October result and
22.64/36.73~pp above SPS-AO/Phoenix. Phoenix is the most relevant seasonal
comparator because it explicitly conditions offloading on sunlight; relative
to it, \sys reduces cost by 65.0\%, latency by 54.6\%, and battery draw by
75.8\%. Its P95 latency and battery draw remain 18.06~s and 20.16~J,
respectively, showing that the frozen controller remains effective under the
alternate illumination pattern, with no seasonal retraining or parameter
retuning.

\noindent\textbf{Impact of constellation.}
At 5~tasks/s, Table~\ref{tab:transfer-summary}(b) changes the topology from
Starlink to the sparser polar OneWeb constellation. With frozen policies,
\sys reaches 86.91\% TAR, 14.54~pp above the strongest baseline, SPS-AO;
LocalDense reaches only 31.68\%, showing the continued need for collaborative
execution under the sparser topology. \sys also achieves the lowest cost,
0.121. Among accomplished tasks, its 8.74~s mean latency is 46.6\% lower than
Phoenix's. SPS-AO reports a lower mean/P95 latency of 1.99/7.20~s, but over a
smaller accomplished-task set. \sys records 3.60/10.18~J mean/P95 battery
draw, lower than all collaborative baselines, while limiting accuracy loss to
1.71\%, compared with 3.00\% for SPS-AO. These results show that its swarm screen-
ing and trajectory planning adapt to changed contact geom-
etry without any constellation-specific retraining.

\begin{table}[t]
\centering
\caption{Generality across seasons and constellations.}
\label{tab:transfer-summary}
\footnotesize
\setlength{\tabcolsep}{3pt}
\resizebox{\columnwidth}{!}{%
\begin{tabular}{lccccc}
\hline
Method & TAR (\%) \(\uparrow\) & $\Phi$ \(\downarrow\) & Mean/P95 Lat. (s) \(\downarrow\) & Mean/P95 Batt. (J) \(\downarrow\) & Acc. loss (\%) \(\downarrow\) \\
\hline
\multicolumn{6}{l}{\emph{(a) Starlink seasonal illumination: October \(\rightarrow\) June}} \\
GroundOnly & 5.19 & 0.381 & 29.19/29.92 & 11.82/28.00 & 0.00 \\
LocalDense & 24.02 & 0.319 & 28.97/29.99 & 3.92/9.97 & 0.00 \\
Phoenix & 54.61 & 0.356 & 14.90/29.89 & 27.64/86.07 & 0.00 \\
MARATD3 & 63.30 & 0.307 & 15.00/25.93 & 20.93/35.75 & 0.00 \\
SPS-AO & 68.70 & 0.206 & 6.08/29.93 & 17.81/20.14 & 2.97 \\
\sys & 91.34 & 0.125 & 6.77/18.06 & 6.69/20.16 & 1.73 \\
\hline
\multicolumn{6}{l}{\emph{(b) OneWeb constellation}} \\
GroundOnly & 7.23 & 0.293 & 29.18/29.92 & 0.149/0.74 & 0.00 \\
LocalDense & 31.68 & 0.308 & 28.90/29.99 & 2.53/2.53 & 0.00 \\
Phoenix & 49.68 & 0.251 & 16.38/29.92 & 11.60/30.35 & 0.00 \\
MARATD3 & 69.21 & 0.363 & 20.60/29.87 & 20.97/34.34 & 0.00 \\
SPS-AO & 72.37 & 0.167 & 1.99/7.20 & 18.00/21.85 & 3.00 \\
\sys & 86.91 & 0.121 & 8.74/27.60 & 3.60/10.18 & 1.71 \\
\hline
\end{tabular}}
\vspace{0.2cm}
\end{table}

\begin{figure}[t]
  \centering
  \includegraphics[width=\columnwidth]{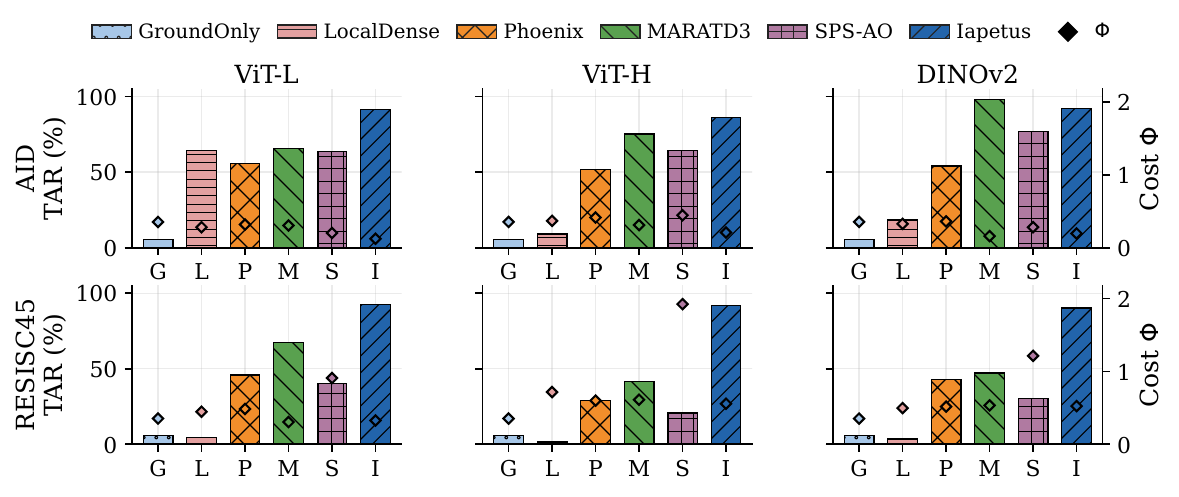}
  \caption{Applicability across models and datasets.}
  \label{fig:model-dataset-transfer}
\end{figure}

\noindent\textbf{Impact of models.}
At 5~tasks/s, the first row of Figure~\ref{fig:model-dataset-transfer} varies
the backbone while retaining AID. Different backbones change both per-layer
computation and intermediate-transfer profiles, altering execution time,
activation traffic, and the savings available from pruning. For fairness,
SPS-AO is trained independently for every model--dataset combination, and
\sys uses the corresponding trained actor and predictor. \sys sustains
86.08--92.28\% TAR, a spread of only 6.20~pp, while keeping cost between 0.122
and 0.208 and below GroundOnly/Phoenix across all three backbones. The heavier
ViT-H workload most directly affects LocalDense, whose TAR falls from 64.14\%
on ViT-L to 9.06\%, and SPS-AO falls to 63.60\%. Although MARATD3 is highest
on DINOv2, \sys leads every baseline on ViT-L and ViT-H and remains
consistently strong across the three backbones.

\noindent\textbf{Impact of datasets.}
At the same rate, the second row replaces AID with RESISC45. Dataset changes
affect both execution demand and content redundancy: under comparable ingress
payloads, RESISC45 incurs greater aggregate inference and codec work, while its
content changes the quality effect of a given pruning control. \sys maintains
90.19--92.51\% TAR and keeps cost between 0.327 and 0.557.
LocalDense falls to 1.63--4.47\% TAR as dense local computation saturates, while MARATD3 drops to 41.79\% on ViT-H and 47.22\% on DINOv2. SPS-AO reaches
only 20.69--40.44\% because its fixed three-cut pipeline incurs greater codec
and execution overhead. Across the three backbones, \sys leads the strongest
baseline by 25.20--50.23~pp, showing consistent adaptation across datasets
under changing workloads.

\vspace{-0.2cm}
\subsection{Runtime Overhead and Scalability}
\label{subsec:overhead}

We decompose decision-inclusive latency into scheduling and post-commit
execution, \(D^{\mathrm{all}}=D^{\mathrm{sch}}+D^{\mathrm{exe}}\), and evaluate
both separately. To quantify scheduling overhead, we time the
frozen ViT-L planning pipeline on Jetson AGX Orin in MAXN mode.
To validate the trace-driven execution model, we replay 600~s of the
committed rate-5 task stream under matched HIL conditions. We report mean
absolute percentage error (MAPE) for execution latency
and onboard battery draw.
Separately, we scale the constellation from 24 to 48 orbital planes under the
default candidate budget
$(K_{\mathrm{swarm}},K_{\mathrm{cand}})=(8,4)$ and vary this budget at 24 planes.

\begin{table}[t]
\centering
\caption{Scheduling latency and HIL fidelity.}
\label{tab:scheduling-fidelity}
\footnotesize
\textit{(a) Scheduling latency by stage.}\\[-2pt]
\renewcommand{\arraystretch}{1.0}
\setlength{\tabcolsep}{2.8pt}
\resizebox{\columnwidth}{!}{%
\begin{tabular}{lcccc}
\hline
 & Cand.\ gen. & Plan.\ (max) & Sel.\ \& commit & Total \\
\hline
Mean (ms) \(\downarrow\) & 49.58 & 167.87 & 0.045 & 217.50 \\
P95 (ms) \(\downarrow\) & 51.77 & 211.10 & 0.050 & 262.91 \\
\hline
\end{tabular}%
}
\par\vspace{1pt}
\textit{(b) HIL execution fidelity at 5~tasks/s.}\\[-2pt]
\newcommand{\hilfidelitytablestretch}{0.85}
\newcommand{\hilfidelitytablevspace}{0pt}
\newsavebox{\hilfidelitytablebox}
\newsavebox{\hilfidelityplotbox}
\sbox{\hilfidelityplotbox}{%
  \includegraphics[width=0.47\columnwidth]{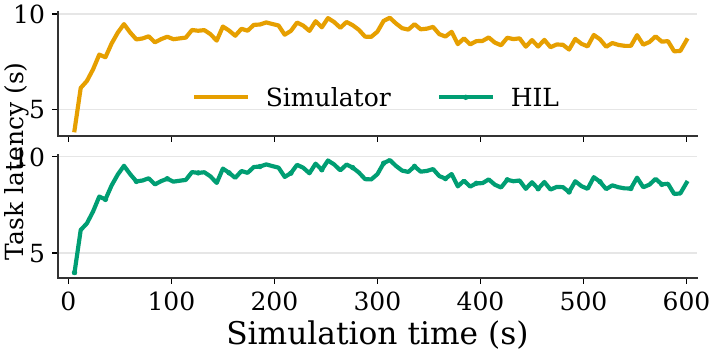}%
}%
\sbox{\hilfidelitytablebox}{%
  \setlength{\tabcolsep}{3pt}%
  \renewcommand{\arraystretch}{\hilfidelitytablestretch}%
  \begin{tabular}{@{}l|>{\raggedleft\arraybackslash}r@{}}
    Metric & Value \\
    \hline
    HIL exec.\ latency & 8.78~s \\
    Latency MAPE & 1.06\% \\
    HIL battery draw & 10.10~J \\
    Battery-draw MAPE & 7.64\% \\
    HIL transfer share & 97.23\% \\
    \hline
  \end{tabular}%
}%
\noindent\begin{minipage}[t]{0.58\columnwidth}
  \vspace{0pt}\usebox{\hilfidelityplotbox}
\end{minipage}\hfill
\begin{minipage}[t]{0.40\columnwidth}
  \vspace{\hilfidelitytablevspace}\usebox{\hilfidelitytablebox}
\end{minipage}
\vspace{0.2cm}
\end{table}

\noindent\textbf{Scheduling and execution latency.}
Table~\ref{tab:scheduling-fidelity}(a) shows that candidate generation and
parallel per-candidate planning dominate scheduling latency; selection and
commit add only 0.045~ms on average. The total is 217.50~ms mean and 262.91~ms
at P95. At rate 5, source candidate generation averages 49.58~ms, below
the 200~ms between arrivals, and overlaps with leader planning for earlier
tasks. Because scheduling latency is only 0.73\% of the default 30-s
deadline, we omit it from the end-to-end latency reported by the
large-scale experiments.

\noindent\textbf{HIL execution fidelity.}
Table~\ref{tab:scheduling-fidelity}(b) shows that modeled latency closely
tracks physical execution, with 1.06\% MAPE.
The HIL runs average 10.10~J battery draw with 7.64\% MAPE. Activation transfer
accounts for 97.23\% of HIL execution time, closely matching its 97.3\% share
in the rate-5 main experiment (6.79~s of 6.98~s). This agreement confirms that
the trace-driven model accurately captures the communication component that
dominates end-to-end latency, beyond matching aggregate latency and battery draw.

\begin{table}[t]
\centering
\caption{Scheduling scalability (latency: Mean/P95).}
\label{tab:scheduling-scalability}
\scriptsize
\setlength{\tabcolsep}{3.5pt}
\textit{(a) Scaling with constellation size.}\\[-2pt]
\resizebox{\columnwidth}{!}{%
\begin{tabular}{ccccc}
\hline
Planes & Summary (B) \(\downarrow\) & Sats. &
Cand. gen. (ms) \(\downarrow\) & Planning (ms) \(\downarrow\) \\
\hline
24 & 596 & 40 & 49.58/51.77 & 217.49/262.87 \\
36 & 597 & 40 & 87.40/89.70 & 263.53/323.02 \\
48 & 601 & 40 & 132.22/134.36 & 306.90/365.61 \\
\hline
\end{tabular}}
\par
\textit{(b) Candidate budgets at 24 planes.}\\[-2pt]
\resizebox{0.88\columnwidth}{!}{%
\begin{tabular}{ccccccc}
\hline
$K_s$ & $K_c$ & Summary (B) \(\downarrow\) & Swarms & Satellites & Plans & Planning (ms) \(\downarrow\) \\
\hline
4 & 1 & 297 & 4 & 20 & 1 & 208.19/225.61 \\
8 & 4 & 596 & 8 & 40 & 4 & 217.49/262.87 \\
16 & 8 & 1193 & 16 & 80 & 8 & 251.62/325.58 \\
\hline
\end{tabular}}
\vspace{0.2cm}
\end{table}

\noindent\textbf{Constellation scaling.}
Scaling from 24 to 48 planes doubles the constellation from 528 to 1,056
satellites. Table~\ref{tab:scheduling-scalability}(a) shows that source-side
candidate generation grows with the topology, while the parallel leader path
remains bounded by the fixed candidate budget. Mean planning latency increases
from 217.49 to 306.90~ms, and detailed planning remains limited to 40
satellites and 596--601 summary bytes per task. The byte count covers the
compact serialized swarm summaries used for candidate screening. Thus,
constellation growth primarily
affects source-side screening.

\noindent\textbf{Candidate budget.}
Table~\ref{tab:scheduling-scalability}(b) shows that parallel leader planning
uses the maximum rather than the sum of the retained candidates' latencies.
Increasing the budget from $(4,1)$ to $(16,8)$ raises the mean from 208.19 to
251.62~ms. Together with \cref{subsec:candidate-reduction}, this supports the
default $(8,4)$ budget, which retains the exhaustive best candidate for 97.5\%
of tasks with a 1.66\% P95 score gap while bounding planning latency.

\section{Related Work}
\label{sec:related_work}

\noindent\textbf{Token-efficient ViT inference.}
Token-efficient ViTs shorten visual sequences through token pruning, merging,
or adaptive sampling. Existing methods use learned importance, token
similarity, or input-dependent policies to discard or aggregate redundant
tokens across Transformer blocks, thereby reducing subsequent attention
computation~\cite{rao2021dynamicvit,yin2022avit,fayyaz2022ats,
liang2022evit,bolya2023token}. Recent vision--language and video systems
further reduce visual tokens after encoding or exploit redundancy across
frames~\cite{chen2024fastv,yang2025visionzip,zhang2025sparsevlm,
tao2025dycoke}. These techniques optimize token efficiency within the model,
but do not jointly determine how content-aware token compression should
interact with distributed layer execution. \sys instead exposes pruning as a
runtime decision whose quality and payload consequences are considered jointly
with layer offloading.

\noindent\textbf{Collaborative Transformer inference in MEC.}
Edge systems distribute Transformer inference through pipeline, tensor,
sequence, and data parallelism, which expose different computation and
communication trade-offs~\cite{hu2024voltage,lee2024hepti,hu2022pipeedge,
deng2024daca,xiong2024epipe,xiong2026mp3}. ViT-specific systems further
decompose models across devices or combine split inference with token compression
to reduce computation and intermediate traffic
~\cite{xu2024devit,jiang2025janus,liu2025lvmscissor,liang2026mercury}.
Most assume terrestrial device--edge settings with relatively stable
participants and edge paths. SEC instead couples time-varying ISL/GSL contacts
with sunlight-dependent energy and finite batteries. Collaborative inference
must therefore be tailored to these contact and energy dynamics.

\noindent\textbf{Collaborative inference and scheduling in SEC.}
SEC systems optimize task offloading or CNN partitioning across
satellites and ground resources under dynamic connectivity and energy
constraints~\cite{xu2023coinleo,chen2025slice,peng2026aptsat,
chen2026hierarchical}. These formulations do not capture a ViT-specific
communication bottleneck: unlike CNN feature maps that shrink with depth, ViT
activations can remain comparable to or larger than the raw input, making
transfer a first-order partitioning cost. Recent work
establishes the viability of Transformer inference in SEC through compressed
submodel chains or topology-aware patch and token scheduling
~\cite{zhang2026communication,lei2026adpssat}. They do not, however, jointly
adapt content-aware token compression and layer offloading across depth.
Building on these efforts, \sys plans these decisions as a trajectory
under finite contacts and coordinates complete plans across swarms.

\section{Conclusion and Future Work}
\label{sec:conclusion}

We present \sys, a content-aware collaborative ViT inference system for
satellite edge computing. \sys formulates each decision as a complete token
compression and layer offloading trajectory, capturing how pruning reshapes
subsequent computation and transfer payloads under dynamic contacts, workload,
and battery conditions. Its hierarchical scheduler bounds constellation-wide
planning and constructs feasible trajectories online from detailed local state
using content-aware prediction and hybrid planning. HIL validation confirms
execution fidelity,
while constellation replay across ViT workloads and constellation settings
shows strong task accomplishment with favorable end-to-end latency, onboard battery draw, and task quality.

Future work will extend \sys in three directions. First, failure-aware
execution can augment one-time commitment with backup trajectories, checkpoint
recovery, and runtime migration when unexpected node or link failures invalidate
the selected plan. Second, multi-source coordination can address concurrent and
bursty EO arrivals that compete for shared swarms, ground links, and resource
reservations. Finally, emerging architectures, including sparse
mixture-of-experts and state-space models, expose new input-dependent execution
structures. Supporting them will require extending the current content-aware
abstraction to model-specific compression, placement, and routing decisions.

\ifdefined\ARXIVVERSION
\else
  \clearpage
\fi

\bibliographystyle{ACM-Reference-Format}
\bibliography{reference}

\clearpage

\appendix

\section{Lyapunov Derivation and Approximation Residual}
\label{sec:appendix-lyapunov}

This appendix derives the plan terms used in
Eq.~\eqref{eq:plan-dependent-phi}. For
\(\Theta(t)=\{Q_s(t),Z_s(t):s\in\mathcal S\}\), define
\[
L(\Theta(t))=
\frac{1}{2}\sum_{s\in\mathcal S}
\left(\frac{Q_s(t)-\theta_s}{Q_{\mathrm{ref}}}\right)^2
+\frac{1}{2}\sum_{s\in\mathcal S}Z_s^2(t).
\]
Write the workload admitted to swarm \(s\) in slot \(t\) as
\[
W_s(t)=\sum_{k\in\mathcal A_t}
\mathbf 1[s_k=s]W_{k,s}(q_k).
\]
For the queue update in Eq.~\eqref{eq:queue-dynamics}, the standard quadratic
inequality gives
\[
\begin{aligned}
&\frac{(Q_s(t+1)-\theta_s)^2-(Q_s(t)-\theta_s)^2}
{2(Q_{\mathrm{ref}})^2}\\
&\quad\le
\frac{\Delta^2\mu_s^2(t)+W_s^2(t)}{2(Q_{\mathrm{ref}})^2}
+\frac{Q_s(t)-\theta_s}{(Q_{\mathrm{ref}})^2}
\bigl(W_s(t)-\Delta\mu_s(t)\bigr).
\end{aligned}
\]

Let \(d_s(t)=[B_s(t)-B_s(t+1)]/B_{\mathrm{ref}}\). Platform draw and solar
input are independent of the current assignment. The component attributable
to assignments is bounded by the admitted mean satellite battery draw:
\[
d_s(t)\le d_s^0(t)+
\sum_{k\in\mathcal A_t}\mathbf 1[s_k=s]
\frac{E_{k,s}(q_k)}{B_{\mathrm{ref}}}.
\]
Applying \(([z+d]^+)^2\le z^2+d^2+2zd\) to
Eq.~\eqref{eq:deficit-update} yields
\[
\frac{Z_s^2(t+1)-Z_s^2(t)}{2}
\le \frac{d_s^2(t)}{2}+Z_s(t)d_s(t).
\]
Define the conditional drift over one slot
\[
\Delta(t)=\mathbb E[L(\Theta(t+1))-L(\Theta(t))\mid\Theta(t)].
\]
Assuming bounded arrivals, service, and battery variation in each slot, the
quadratic terms above are bounded by a finite constant \(C\). Summing over
swarms and adding the weighted task execution cost gives
\par
{\scriptsize
\compactdisplayskips%
\[
\Delta(t)+V\mathbb E[\Phi(t)\mid\Theta(t)]
\le C+\Gamma(t)+\mathbb E\!\left[\sum_{k\in\mathcal A_t}
\Psi_{k,s_k}(q_k;t)\mid\Theta(t)\right],
\]
}%
where \(\Gamma(t)\) contains service and background energy terms independent
of the current assignments. Minimizing the part affected by the current
decision produces Eq.~\eqref{eq:plan-dependent-phi}: its first two terms impose
workload and battery penalties under the accumulated pressures, while the last
carries the task execution cost.

Let \(\varepsilon_t\ge0\) collect the residual from swarm screening,
candidate truncation, coarse estimation, learned trajectory search, and
sequential commitment. Assuming
\(\varepsilon_t\le\varepsilon_{\max}<\infty\), the drift bound gains an
additive bounded term. The deployed search is therefore interpreted as an
approximation to the common Lyapunov objective, not as an exact optimizer for
every slot.

For the C3 diagnostic in \cref{subsec:candidate-reduction}, let
\(\Psi_k^\star\) be the minimum detailed score over a tractable exhaustive
task set and \(\Psi_k^{\mathrm{sel}}\) the score selected by the deployed
bounded search. We report the normalized gap
\[
g_k=\frac{\Psi_k^{\mathrm{sel}}-\Psi_k^\star}
{\max(|\Psi_k^\star|,10^{-12})},
\]
together with the fraction of tasks for which the candidate attaining
\(\Psi_k^\star\) survives Top-\(K\) retention. These diagnostics measure search
approximation.

\section{Additional Evaluation Results}
\label{sec:appendix-evaluation}

\begin{table}[t]
  \centering
  \caption{Default system and training parameters.}
  \label{tab:appendix-system-parameters}
  \footnotesize
  \setlength{\tabcolsep}{4pt}
  \renewcommand{\arraystretch}{1.08}
  \textit{(a) Physical and runtime parameters.}\\[-2pt]
  \begin{tabularx}{\columnwidth}{@{}>{\raggedright\arraybackslash}X
      >{\raggedright\arraybackslash}X@{}}
    \toprule
    Parameter & Value \\
    \midrule
    Starlink size (planes $\times$ sats) & $24{\times}22$ \\
    OneWeb size (planes $\times$ sats) & $14{\times}36$ \\
    TLE epoch (October / June) & 2025-10-16 / 2025-06-21 10:00 \\
    Trace horizon / slot duration & 6000 / 0.1~s \\
    \cy{Task input composition} & \cy{32 AID / 176 RESISC45 images
    ($\approx 0.277$~Gbit, 8-bit RGB)} \\
    GSL / ISL bandwidth & 0.1 / $\mathcal U[0.1,1]$~Gbps \\
    GSL / ISL transmit power & 10 / 8.5~W \\
    Baseline / solar input power & 6.62 / 19.2~W \\
    Orin mode (sunlight / eclipse) & MAXN / 30~W \\
    Incremental compute power & 16.42 / 5.59~W \\
    Lyapunov weight $V$ & 50 \\
    Swarm composition & One leader and four members \\
    Queue and service model & Non-preemptive processor and link ready times \\
    \bottomrule
  \end{tabularx}
  \vspace{0.5em}

  \textit{(b) Offline predictor and policy training.}\\[-2pt]
  \begin{tabularx}{\columnwidth}{@{}>{\raggedright\arraybackslash}X
      >{\raggedright\arraybackslash}X@{}}
    \toprule
    Parameter & Value \\
    \midrule
    Predictor / input descriptor & Histogram GBT / 16 statistics \\
    Isolated predictor records & $1{,}134{,}000$ \\
    History-conditioned records & $48{,}000$ \\
    GBT loss / depth / iterations & Squared error / 6 / 120 \\
    GBT learning rate & 0.08 \\
    GBT $L_2$ regularization & 1.0 \\
    Actor and critic networks & Two 256-unit hidden layers \\
    SAC learning rate / batch size & $3{\times}10^{-4}$ / 256 \\
    Replay capacity / initial transitions & $200{,}000$ / $20{,}000$ \\
    Discount $\gamma$ / target update $\tau$ & 1.0 / 0.005 \\
    Target entropy (pruning / node) & $-1$ / $0.5\ln 6$ \\
    Offline updates & $3{,}000$ \\
    \cy{Training episodes / episode length} & $720$ / $64$ steps \\
    \bottomrule
  \end{tabularx}
\end{table}

\noindent\textbf{Admission under link fluctuations.}
Table~\ref{tab:appendix-link-admission} decomposes TAR into the fraction of
released tasks admitted for execution and the fraction of admitted tasks
accomplished within both the deadline and accuracy requirements.

\begin{table}[t]
  \centering
  \caption{Admission decomposition under link fluctuations at 5 tasks/s.}
  \label{tab:appendix-link-admission}
  \footnotesize
  \setlength{\tabcolsep}{4.2pt}
  \resizebox{\columnwidth}{!}{%
  \begin{tabular}{lccc}
    \hline
    Method & Admitted / released (\%) & Accomplished / admitted (\%) & TAR (\%) $\uparrow$ \\
    \hline
    GroundOnly & 4.2 & 97.6 & 4.1 \\
    LocalDense & 24.7 & 99.8 & 24.7 \\
    Phoenix & 29.4 & 100.0 & 29.4 \\
    MARATD3 & 56.4 & 100.0 & 56.4 \\
    SPS-AO & 78.1 & 72.7 & 56.8 \\
    \sys & 72.0 & 93.3 & 67.2 \\
    \hline
  \end{tabular}}
\end{table}

\begin{figure}[t]
  \centering
  \includegraphics[width=\columnwidth]{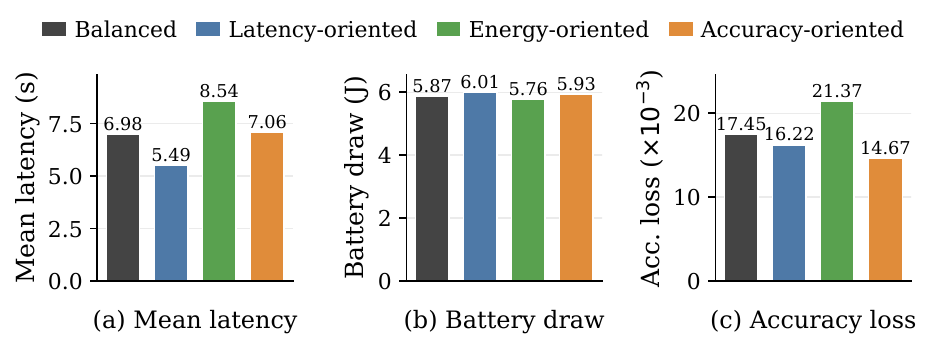}
  \caption{Objective-weight sensitivity.}
  \label{fig:appendix-objective-weight-sensitivity}
\end{figure}

\noindent\textbf{Sensitivity to objective weights.}
We compare four profiles over $(\omega_D,\omega_E,\omega_A)$: Balanced
$(0.3,0.3,0.4)$, Latency-oriented $(0.6,0.2,0.2)$, Energy-oriented
$(0.2,0.6,0.2)$, and Accuracy-oriented $(0.2,0.2,0.6)$. All profiles use the
same AID+ViT-L ToMe configuration at 5 tasks/s and the same evaluation seeds.
All profiles use the same frozen actor, with the objective weights supplied as inputs.

Figure~\ref{fig:appendix-objective-weight-sensitivity} shows that each
oriented profile improves its target metric relative to Balanced: latency
weighting reduces mean latency by 21.3\%, energy weighting reduces mean battery
draw by 1.1\%, and accuracy weighting lowers mean accuracy loss by 16.0\%.
These directions hold for every seed. These gains introduce corresponding
tradeoffs. The Energy-oriented profile loses 2.32~pp of TAR and raises mean
latency and accuracy loss by 22.3\% and 22.5\%, while the Latency-oriented
profile draws 1.6\% more battery. Balanced retains competitive TAR without such one-sided
degradation and is therefore used as the default joint objective.

\end{document}